\documentclass[preprintnumbers,amsmath,amssymb,floatfix,11pt,prd,onecolumn,superscriptaddress,nofootinbib]{revtex4-1}

\usepackage{graphicx}
\usepackage{dcolumn}
\usepackage{amsmath,amssymb,mathtools}
\usepackage{epstopdf}
\usepackage{verbatim}
\usepackage{amsmath}
\usepackage[colorlinks=true, citecolor=blue, urlcolor = blue, linkcolor= blue,
bookmarks=true]{hyperref}
\labelformat{section}{Section #1}
\labelformat{subsection}{Section #1}
\labelformat{subsubsection}{Section #1}
\labelformat{subsubsubsection}{Section #1}
\labelformat{equation}{Eq.~(#1)}
\labelformat{figure}{Fig.~#1}
\labelformat{subfigure}{Fig.~\thefigure#1}
\labelformat{table}{Tab.~#1}
\labelformat{appendix}{Appendix #1}

\begin{document}

\title{Strong gravitational lensing of a five-dimensional charged, equally rotating black hole with a cosmological constant}

\author{Md Sabir Ali}
\email{alimd.sabir3@gmail.com}
\affiliation{Key Laboratory of Quantum Theory and Applications of Ministry of Education, Lanzhou Center for Theoretical Physics, Lanzhou University, Lanzhou 730000, China}
\affiliation{{Key Laboratory of Theoretical Physics of Gansu Province, Institute of Theoretical Physics $\&$ Research Center of Gravitation, Lanzhou University, Lanzhou 730000, China}}
\affiliation{School of Physical Science and Technology, Lanzhou University, Lanzhou 730000, China}
\affiliation{Department of Physics, Mahishadal Raj College, West Bengal 721628, India}
\author{Shagun Kaushal}
\email{shagun123@iitd.ac.in}
\affiliation{Department of Physics, Indian Institute of Technology Delhi, Hauz Khas, New Delhi-110016, India}
%%%%

\author{Yu-Xiao Liu}
\email{liuyx@lzu.edu.cn, the corresponding author}
\affiliation{Key Laboratory of Quantum Theory and Applications of Ministry of Education, Lanzhou Center for Theoretical Physics, Lanzhou University, Lanzhou 730000, China}
\affiliation{{Key Laboratory of Theoretical Physics of Gansu Province, Institute of Theoretical Physics $\&$ Research Center of Gravitation, Lanzhou University, Lanzhou 730000, China}}
\affiliation{School of Physical Science and Technology, Lanzhou University, Lanzhou 730000, China}

\begin{abstract}
We study the lensing phenomena of the strong gravity regime of five-dimensional charged, equally rotating black holes with a cosmological constant, familiarly known as the Cveti\v c-L\"u-Pope black holes. These black holes are characterized by three observable parameters, the mass $M$, the charge $Q$ and the angular momentum $J$, in addition to the cosmological constant. We investigate the strong gravitational lensing observables, mainly the photon sphere radius, the minimum impact parameter, the deflection angle, the angular size, and the magnification of the relativistic images. We model the $M87$ and $SgrA^*$ for these observables. We also focus on the relativistic time delay effect in the strong-field regime of gravity and the impact of the observable on it. The analytical expressions for the observables of the relativistic images with vanishing angular momentum ($j=0$) are discussed in some detail. We shed a light on the gravitational time delay effect by incorporating the lensing observables. The gravitational time delay has a direct consequence on the photon sphere radius and hence on the quasinormal modes.
\end{abstract}
\maketitle

\section{Introduction}
The recent year witnessed a flurry of activities regarding the astronomical measures of the general relativistic effects, especially the observation of the black hole shadows, e.g., for the M87 and SgA*, and the gravitational wave data from the LIGO scientific/Virgo collaborations \cite{LIGOScientific:2017ycc, LIGOScientific:2016aoc, LIGOScientific:2018dkp,  LIGOScientific:2019fpa, Wielgus:2022heh, KAGRA:2023pio}. The data from the Event Horizon Telescope (EHT) regarding the shadows of black holes confirmed the validity of Einstein's general relativity (GR) and established it as a firm base for any general relativistic theory of gravity \cite{Akiyama:2019cqa, Akiyama:2019brx, Akiyama:2019sww, Akiyama:2019bqs, Akiyama:2019fyp, Akiyama:2019eap}. As is known from its definition, gravitational lensing is the deflection of light rays when they pass close to black holes \cite{Darwin, Frittelli:1999yf, Eiroa:2010wm}. Gravitational lensing is utilized as an important tool in astrophysics; for example, it is used to measure the mass of compact objects. The formation of Einstein's ring is one of the remarkable observations made during the lensing phenomenon. The ring forms when the bright source (the star), the black hole (the lens), and the observer are aligned in a perfect way. Gravitational lensing has been investigated in detail for a variety of black hole solutions in GR as well as in other theories of gravity. For reviews on both weak and strong gravitational lensings in both equatorial ($\theta=\pi/2$) and non-equatorial ($\theta\neq \pi/2$) planes, we refer our readers to~\cite{Bozza:2001xd, Bozza:2002zj, Bozza:2007gt, Bozza:2002af, Kuang:2022xjp, Sengo:2022jif, Bartelmann:2016dvf, Hoekstra:2008db} and references therein. The other analytical studies black holes have been reviewed and highlighted in general relativity as well as in the modified scenarios \cite{AbhishekChowdhuri:2023ekr, Ghosh:2022mka}. The connection of the shadow and lensing properties has been explored in the context of modified gravity theories \cite{EslamPanah:2020hoj, Hendi:2022qgi}
%%%%%%%%%%%%%%%%%%%%%%%%%%%%%%%%%%%%%%%%%%%%%%%%%%%%%%%%%%%%%%%%%%%%%%%%%%%%%%%%%%%

From the strong lensing point of view, the light rays pass close enough to the black hole wind once or several times before leaving that region and reaching the observer. The idea that black holes act as gravitational lenses was first put forward by Darwin \cite{Darwin} and subsequently elaborated in Refs.~\cite{Ohanian:1987pc, Luminet:1979nyg}. The most recent theoretical investigation of strong field gravitational lensing is mainly attributed to the seminal work of Virbhadra and Ellis \cite{Virbhadra:1999nm, Virbhadra:2002ju, Virbhadra:2008ws}, where they discussed the formation of the ring, its position, and the magnification of relativistic images for the Schwarzschild black hole. Another work discussed the role of the scalar-to-mass ratio \cite{Virbhadra:1998dy} in the lensing context. After this formulation, Frittelli, Killing and Newman obtained more rigorous analytical descriptions of the exact lens equation and the integral expressions for its solution \cite{Frittelli:1999yf}. A comparison of their results with those given by Virbhadra and Ellis was also presented. Later, Bozza \cite{Bozza:2001xd, Bozza:2002zj, Bozza:2007gt, Bozza:2002af} and then Tsukamoto \cite{Tsukamoto:2016jzh, Tsukamoto:2017fxq} proposed a completely new method for the study of strong lensing for a generic spherically symmetric static spacetime. These methods are widely accepted by the scientific community and are still used as applications in the study of strong lensing phenomena. These methods were applied to study the various black hole systems, including solutions in various modified theories of gravity \cite{Islam:2021dyk, Ghosh:2020spb, Whisker:2004gq, Abbas:2021whh, Eiroa:2005ag, Gyulchev:2006zg, Ghosh:2010uw, Gyulchev:2012ty, Molla:2023hog, Grespan:2023cpa, Kumar:2022fqo, KumarWalia:2022ddq, Kumar:2021cyl, Lu:2021htd, Ali:2021psk, Soares:2023err, Hsieh:2021scb, Soares:2023uup}. More recently, the modeling of the supermassive compact object M87$^*$ as Schwazrschild black hole and subsequent analysis of the variations of the three different magnified images of different orders has been explored in the observational context \cite{Virbhadra:2022iiy}. These investigations may result in many outcomes that may be important for the operation of the generation Event Horizon Telescope (ngEHT). Further, the ratios of the mass to distance, and the lens-source to observer-source are used for the measure of the compactness of the black hole lenses \cite{Virbhadra:2022ybp}.\\
%%%%%%%%%%%%%%%%%%%%%%%%%%%%%%%%%%%%%%%%%%%%%%%%%%%%%%%%%%%%%%%%%%%%%%%%%%%%%%%%%%%%%%%%%%
We take one presumption regarding the validity of the four-dimensional Ohanian lens equation in a five-dimensional scenario, we shed light into the issues concerned with the extra dimensions, the validity of the four-dimensional lens equations in a dimension $4+n$, where $n$ is the number of the extra spatial dimensions. As the ratios of mass to radius for the planets in our solar system are quite small, the corrections to the gravitational lensing by planetary objects in the solar system are negligibly small and hence beyond the reach for near future experiments \cite{He:1999fe}. Moreover, the  gravitational lensing by supermassive black holes with known masses and radii there will be of potential effects of the extra dimensions in larger amount. The future precision experiments from LIGO/Virgo or the Event Horizon Telescope (EHT) or ngEHT on the strong gravitational lensing for the supermassive black holes can provide important information about the possible viable theory of gravitation apart from Einstein GR and hopefully the extra dimensions \cite{Nandi:2024map, Zahid:2024nvx, Shavelle:2024vwt, Jiang:2023img, Ayzenberg:2023hfw}. There are many modified theories of gravity, such as tensor-scalar theories or theories with extra spatial dimensions apart from usual four spacetime dimensions. It is legitimate to speculate the ideas about those alternative gravity theories which may be consistent with precision tests using observational data and find an ultimate theory of gravity. There may arise different type of gravitational interactions in the modified theories in addition to the existing interactions in Einstein general relativity counted as the correction terms. We do hope that the experimental measurements can put much strong constraints for other theories or even rule out some of the theories. In this paper we study gravitational lensing in theories with extra dimensions along the same line of thought that are outlined above. As a result, we can observe in the various observable parameters such as the deflection angle, the angular position, the angular separation, the image magnifications, and also the relativistic time delay effects. All of these parameters are highly affected due to introduction of the extra dimensions.\\

The string theory is a promising candidate for the unified theory, where gravity is unified with other forces of nature. The prediction of extra dimensions and its relation to the physical observable attracted a lot of attention because it borrowed the signature of the string and hence made string theory a correct one. 
The search for extra dimensions had been a challenging task and a long standing problem. It always has been an active area of investigation after the use of quasinormal modes (QNMs) arising from the perturbed black holes spacetimes in higher-dimensions. It may also be made possible through the analysis of the spectrum of the Hawking radiation. Extra dimensions could be detected through the investigation of QNMs spectra of higher-dimensional spacetimes \cite{Shen:2006pa, Harris:2005jx, Casals:2005sa, Creek:2006ia, Kanti:2005xa, Kanti:2006ua, Creek:2006je, Konoplya:2011qq, Kodama:2009rq, Konoplya:2008rq, Nozawa:2008wf, Chen:2007ay, Chen:2007pu}, which may be probed via the gravitational wave detectors in the near future. One of the possible insights may be useful when one studies the higher-dimensional black holes and black branes. Additionally, the correlation between the strong gravitational lensing and QNMs in the high momentum or eikonal approximation could have been probed if careful investigation had been carried out. In this context, strong gravitational lensing is an effective method for investigating the presence of dimensions higher than four.
 %Therefore, gravitational lensing can also be used to probe the extra dimensions. In this direction, the study of strong lensing in higher-dimensional black hole solutions is worth investigating. 
The search for extra dimensions was probed through the gravitational lensing of a charged-neutral/charged Kaluza-Klein squashed black hole \cite{Liu:2010wh,Sadeghi:2012bj,Ji:2013xua}, the squashed Kaluza-Klein G\"odel black hole for both charged and charged-neutral cases \cite{Chen:2011ef,Sadeghi:2013ssa}, the black hole with extra-dimensions and the Kalb-Ramond field \cite{Majumdar:2006zza,Chakraborty:2016lxo}. These studies help us to understand the effective length scale on which the extra dimension lies. Motivated by these investigations, in our present paper, we wish to study the strong gravitational lensing of the five-dimensional charged equally rotating black hole in the presence of a negative cosmological constant as given in Ref.~\cite{Markeviciute:2018cqs}, when the charged scalar hair has vanishing limit. This black hole solution was originally proposed by Cveti\v c, L\"u and Pope \cite{Cvetic:2004hs}, hereby denoted as the CLP black hole. Such a solution is considered as a coupling of the Einstein-Maxwell system that arises as the bosonic sector of the minimal-gauged five-dimensional supergravity theories. As for the special cases, such a solution reduces all the previously known cases. The solution is characterized by the mass, the charge, the angular momentum, and the cosmological constant. We shall see how the charge, the rotation parameter, and other defining parameters affect the size of the horizon, the photon sphere radius, the formation of relativistic images, the deflection angle, and subsequently other physical observables in the strong field lensing of gravity. On the other hand, in relation to the AdS/CFT correspondence if one is allowed to evaluate the retarded two point functions (response functions) in a gauge covariant theory, it could be made possible by studying their fluctuations around any asymptotically AdS black holes. The poles of these response functions are the characteristic features of the QNMs frequencies of the AdS black hole spacetimes. The black hole spacetimes in higher dimensions are very important in the context of string theory and its subsequent analysis through the AdS/CFT correspondence. Thus, the study of strong gravitational lensing as well as the quasinormal modes in the higher dimensional black holes from the mirror of the AdS/CFT correspondence may help us to extract information about extra dimensions in astronomical observations in future experiments.
\vspace{0.5mm}\\
The main physical quantity that has played the major role in predicting the bending of light beams around black holes in any gravity theory whether it is in the asymptotically flat and AdS spacetime is the photon sphere radius. For the study of the 
the photon sphere in the AdS/CFT realm, our findings may be based on three important streams. Firstly, a symmetry group $SL(2,R)$ indicates the set of QNM frequencies at the photon sphere radius of the Schwarzschild and Kerr black holes spacetimes at the asymptotic flat limits \cite{Raffaelli:2021gzh, Hadar:2022xag}. An extension of such study was landed into the search of the warped geometries and subsequent investigations were carried out \cite{Kapec:2022dvc, Chen:2023zvd, Chen:2022fpl, Fransen:2023eqj}. Secondly, the QNMs frequencies for the Schwarzschild-AdS black hole is a long-standing pathway: the physics has very rich structure in its own right, and the remarkable study in the WKB approximation has been performed for the QNMs analyses of the various AdS black holes \cite{Chan:1996yk, KalyanaRama:1999zj, Horowitz:1999jd}. Thirdly, the photon sphere radius is emerged as a direct consequence of the Einstein ring while imaging the black holes in any gravitational theory, and the transformation of the images on the onset of the holographic CFTs indeed was used to produce the Einstein rings of such types \cite{Hashimoto:2019jmw, Hashimoto:2018okj}.

The organization of the paper is as follows. In Section~\textcolor{blue}{II}, we briefly discuss the solution of the five-dimensional charged equally rotating black hole solution as given by Cveti\v c, L\"u and Pope. We also investigate the horizon structure and parametric bounds on the charge and rotation parameters. Section~\textcolor{blue}{III} is devoted to discussing the formulation of the deflection angle using Bozza's method. We discuss the effective potential, the photon sphere radius, the impact parameter, and their connections. We also discuss them in the limiting cases, when $j=0$. Next, we discuss the lensing observable and the formation of the relativistic images of the black hole in Section~\textcolor{blue}{IV}. Here, we model our concerned black holes as SgrA$^*$ and M87$^*$ to show the effect of the parameters. In Section~\textcolor{blue}{V}, we discuss the time delay effect. Finally, in Section~\textcolor{blue}{VI}, we summarize the paper and conclude our results.

\section{The five-dimensional charged equally rotating black hole solution with a cosmological constant}
We present a brief review of the action and the constituent spacetime before we move on to the main discussion of the lensing phenomena. The expression for the action is given by a consistent truncation and is written as \cite{Markeviciute:2018cqs,Bhattacharyya:2010yg}
\begin{eqnarray}
    \label{action}
    S&=&\frac{1}{16\pi G_5}\int d^5x \sqrt{-g}\Bigg(R+12-\frac{3}{4}F_{\mu\nu}F^{\mu\nu}
     -\frac{3}{8}\left(\left|D_\mu\phi\right|^2-\frac{\partial_\mu(\phi\phi^*)
     \partial^\mu(\phi\phi^*)}{4(4+\phi\phi^*)}-4\phi\phi^*\right) \nonumber\\
    &&+\frac{1}{4\sqrt{-g}}\epsilon^{\alpha\beta\gamma\mu\nu}F_{\alpha\beta}F_{\gamma\mu}A_\nu\Bigg),
    %&=&\frac{N^2}{2\pi^2}\int d^5x \sqrt{-g}\Bigg(R+12-\frac{3}{4}F_{\mu\nu}F^{\mu\nu}-\frac{3}{8}\left(\left|D_\mu\phi\right|^2-\frac{\partial_\mu(\phi\phi^*)\partial^\mu(\phi\phi^*)}{4(4+\phi\phi^*)}-4\phi\phi^*\right)\nonumber\\
%    &&+\frac{1}{4\sqrt{-g}}\epsilon^{\alpha\beta\gamma\mu\nu}F_{\alpha\beta}F_{\gamma\mu}A_\nu\Bigg),
\end{eqnarray}
where
\begin{eqnarray}
    D_\mu\phi =\partial_\mu\phi-2i A_\mu\phi,\;F_{\mu\nu}=\partial_\mu{A_\nu}-\partial_\nu{A_\mu}.
\end{eqnarray}
It is to be mentioned that in \ref{action} we did put the cosmological constant or the curvature radius to be equal to unity. Later, we shall restore it whenever required.
The variation of the action (\textcolor{blue}{1}) with respect to the metric tensor $g_{\mu\nu}$ yields the Einstein's field equations \cite{Markeviciute:2018cqs,Bhattacharyya:2010yg}
\begin{eqnarray}
    \label{EOM}
    R_{\mu\nu}-\frac{1}{2}\left(R+12\right)g_{\mu\nu}
    =-\frac{3}{2}T_{\mu\nu}^{\text{EM}}+\frac{3}{8}T_{\mu\nu}^{{\phi}}.
\end{eqnarray}
The expressions for the energy-momentum tensor of the electromagnetic field $T_{\mu\nu}^{\text{EM}}$, and the scalar field $T_{\mu\nu}^{\phi}$ are
\begin{eqnarray}
    \label{emt}
 T_{\mu\nu}^{\text{EM}} &=& F_{\mu}^\alpha F_{\nu \alpha}-\frac{1}{2}g_{\mu\nu}F_{\alpha\beta}F^{\alpha\beta}, \nonumber\\
 T_{\mu\nu}^{\phi} &=& \frac{1}{2}\Bigg[D_\mu\phi (D_\nu\phi)^*+D_\nu\phi (D_\mu\phi)^*-\frac{1}{2}g_{\mu\nu}\left|D_\alpha\phi\right|^2+2\phi\phi^* g_{\mu\nu}\nonumber\\
  &&-\frac{1}{4(4+\phi\phi^*)}\left(\partial_\mu(\phi\phi^*)\partial_\nu(\phi\phi^*)
 -\frac{1}{2}g_{\mu\nu}\left[\left(\partial_\sigma\phi\phi^*\right)
 \left(\partial^\sigma\phi\phi^*\right)\right]\right)\Bigg].
\end{eqnarray}
The Maxwell and the scalar field equations are given, respectively, as \cite{Markeviciute:2018cqs, Bhattacharyya:2010yg}
\begin{eqnarray}
    \label{maxscalar}
   & \nabla_\nu F_\mu^\nu=\frac{i}{4}\left[\phi(D_{\mu}\phi)^*-\phi^*(D_{\mu}\phi)
     +\frac{1}{4\sqrt{-g}}g_{\mu\sigma}\epsilon^{\sigma\alpha\beta\gamma\nu}
      F_{\alpha\beta}F_{\gamma\nu}\right], \\
 & D_\mu D^\mu{\phi}+\left[\frac{\left[\partial_\sigma(\phi\phi^*)\right]^2}{4(4+\phi\phi^*)^2}
  -\frac{\nabla^2(\phi\phi^*)}{2(4+\phi\phi^*)}\right]\phi=0.
\end{eqnarray}
The stationary, asymptotically anti-de Sitter spacetimes have spherical horizon topology. The generic solutions to such a structure have two independent rotation parameters. To keep the structure simpler we consider the doubly rotating black hole solutions, having the same magnitude of the rotation parameters but with two different orientations. For any generic gauge choice, we can write down the metric $ansatz$ as follows \cite{Markeviciute:2018cqs}:
\begin{eqnarray}
\label{metric}
\mathrm{ds}^2= -f(r)\mathrm{dt}^2+g(r)\mathrm{dr}^2+\Sigma^2(r)\Big[h(r)
 \big(\mathrm{d\psi}+\frac{1}{2}\cos{\theta}\mathrm{d\phi}-\Omega(r)\mathrm{dt}\big)^2+\frac{1}{4}\mathrm{d\Omega}_2^2\Big],
\end{eqnarray}
which is rewritten as
\begin{eqnarray}
\mathrm{ds}^2&=&\Big(-f(r)+\Sigma^2(r) \Omega^2(r)h(r)\Big)\mathrm{dt}^2+g(r)\mathrm{dr}^2+\frac{1}{4}\Sigma^2(r) \mathrm{d\theta}^2-\Sigma^2(r) \Omega(r) h(r) \cos\theta \mathrm{d\phi} \mathrm{dt}\nonumber\\
    &+&\frac{\Sigma^2(r)}{4}\Big( h(r)\cos^2{\theta}+\sin^2{\theta}\Big)\mathrm{d\phi}^2+\Sigma^2(r)h(r)\mathrm{d\psi}^2-2 \Sigma^2(r) \Omega(r)h(r)\mathrm{d\psi}\mathrm{dt} \nonumber\\
    &+&\Sigma^2(r) h(r) \cos{\theta}\mathrm{d\psi}\mathrm{d\phi}.
    \label{metric_2}
\end{eqnarray}
In the above $ansatz$, we can see that the metric on $S^3$ is expressed as a Hopf fibration over the space $\mathbb{C}\mathbb{P}^1$ of unit radius. In this way, the coordinate $\psi$ has the period of $2\pi$ and the $\theta,\;\phi$ coordinates have the coordinate ranges of the usual unit $S^2$ sphere. General five-dimensional, charged, and equally rotating black hole solutions with a cosmological constant were first presented in Ref.~\cite{Cvetic:2004hs}, which is basically a coupled Einstein-Maxwell solution in the bosonic sector of minimal gauged supergravity, and also $\phi=0$ limit of the action (\textcolor{blue}{1}). The black holes are governed by three hairs $\lbrace q, j, m\rbrace$, and an additional fourth parameter, the cosmological constant term. The parameters $\lbrace q, j, m\rbrace$ are related to the conserved
charges $\lbrace Q, J, M\rbrace$. The functions appear in the metric (\textcolor{blue}{8}) have the explicit forms
\begin{eqnarray}
 \Sigma(r) &=& r,~~~ f(r)=\frac{G(r)}{h(r)},~~~ g(r)=\frac{1}{G(r)},  \nonumber \\ h(r)&=&1+j^2\Big(\frac{m}{r^4}-\frac{q^2}{r^6}\Big),~~~
   \Omega(r)=\frac{j}{h(r)}\Big(\frac{m-q}{r^4}-\frac{q^2}{r^6}\Big),\\
 G(r) &=& \frac{1}{r^4}\Big[q^2\Big(1-\frac{j^2}{L^2}\Big)+j^2m\Big]
 -\frac{1}{r^2}\Big[m\Big(1-\frac{j^2}{L^2}\Big)-2q\Big]+\frac{r^2}{L^2}+1. \nonumber
\end{eqnarray}
The angular momentum $J$ and the mass $M$ of the black hole are related to the parameters $\lbrace q, j, m\rbrace$ as follows:
\begin{eqnarray}
    \label{J_M}
    J=\frac{1}{2}j\left(m-q\right),~~~ M=\frac{1}{4}\left[m\left(3+\frac{j^2}{L^2}\right)-6q\right],~~~
    Q=\frac{q}{2}.
\end{eqnarray}
Using the last two relations of the above equations we can express $m$ and $q$ as
\begin{eqnarray}
    \label{MandQ}
m=\frac{2 L^2 (2 M+6 Q)}{j^2+3 L^2},~~~ q=2Q.
\end{eqnarray}
Please note that in the action, we put the AdS radius $L=1$. Here, we restored it for the sake of clarity and calculation. The static and spherically symmetric part of the solution is obtained if one puts
\begin{eqnarray}
    \label{CLP}
\mathrm{ds}^2&=&-f(r)\mathrm{dt}^2+g(r)\mathrm{dr}^2+\Sigma^2(r) \mathrm{d\Omega}_3^2, \\
\mathrm{d\Omega}_3^2&=&(\mathrm{d\psi}+\frac{1}{2}\cos\theta\mathrm{d\phi})^2+\frac{1}{4}\mathrm{d\Omega}_2^2=\mathrm{d\theta}^2+\sin^2\theta \mathrm{d\phi}^2+\cos^2\theta \mathrm{d\psi}^2.
\end{eqnarray}
Obviously, the metric (\textcolor{blue}{12}) is the five-dimensional Reissner-Nordstr$\Ddot{o}$m black holes with a cosmological constant, where $f(r)$ and $g(r)$ are given by \cite{Cvetic:2004hs}
$$f(r)=1+\frac{r^2}{L^2}-\frac{m}{r^2}+\frac{q^2}{r^4},\;\;\;g(r)=1/f(r).$$

The metric possesses three Killing vectors $\partial_{{t}}$, $\partial_{\hat{\phi}}$, and $\partial_{\hat{\psi}}$, where $\hat{\phi}=\psi-\phi$, $\hat{\psi}=\psi+\phi$. For an equally rotating black hole, there are two additional Killing vectors \cite{Frolov:2003en, Frolov:2002xf}
\begin{eqnarray}
 &\cos{\hat{\phi}}{\partial_{\hat{\theta}}}
  -\cot{\hat{\theta}}\sin{\hat{\phi}}{\partial_{\hat{\phi}}}
  +\frac{\sin{\hat{\phi}}}{\sin{\hat{\theta}}}\partial_{\hat{\psi}}, \\
  &-\sin{\hat{\phi}}{\partial_{\hat{\theta}}}
  -\cot{\hat{\theta}}\cos{\hat{\phi}}{\partial_{\hat{\phi}}}
  +\frac{\cos{\hat{\phi}}}{\sin{\hat{\theta}}}\partial_{\hat{\psi}},
\end{eqnarray}
where $\hat{\theta}=2\theta$.

\begin{figure}[h]
    \includegraphics[scale=0.6]{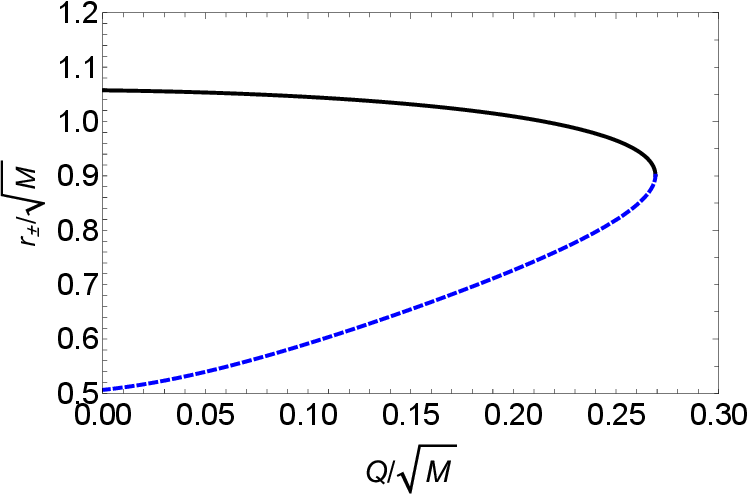}\hspace{0.1cm}
    \includegraphics[scale=0.6]{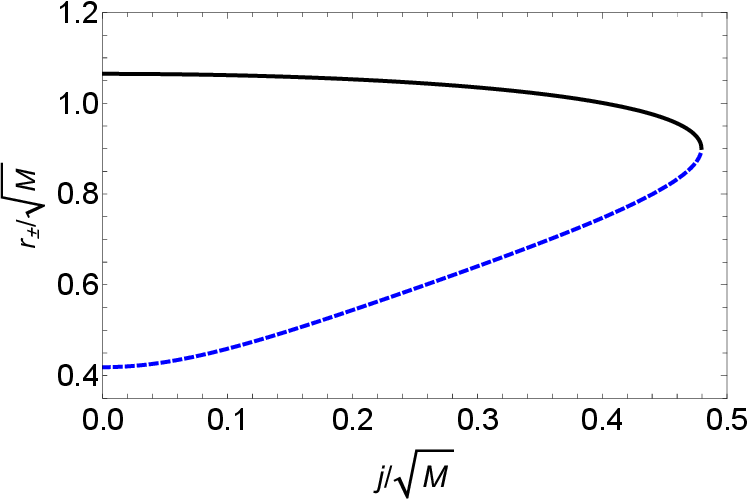}
    \caption{The plot of the Cauchy horizon (blue dashed line) and the event horizon (black solid lines) vs the charge parameter $Q$ for $j=0.1, L=10$ (the left plot) and the rotation parameter $j$ for $Q=0.1, L=10$  (the right plot).}
    \label{fig:horizon}
\end{figure}

\begin{figure}
    \centering
    \includegraphics[scale=0.6]{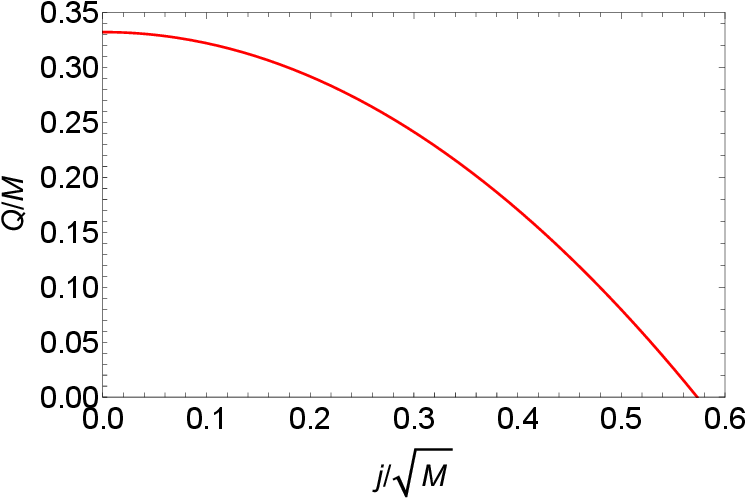}
    \caption{The plot of the black hole charge parameter $Q/M$ vs the rotation parameter $j/\sqrt{M}$. This is the parameter space in the plane $(Q/M,j/\sqrt{M})$ for the CLP black hole. The red line represents the extremal black hole with degenerate Cauchy and the event horizons. The region above the red line is the no black hole region while the region below it corresponds to the black hole region.}
    \label{fig:Plot1}
\end{figure}

The event horizon is a well-defined boundary that is a null hypersurface and it comprises the outward null geodesics, which are not capable of hitting the null infinity in the future. The event horizon is a solution of $F(r)=0$:
$$r_h^6+L^2 r_h^4+r_h^2 \left(j^2 m-L^2 m+2 L^2 q\right)+j^2 L^2 m-j^2 q^2+L^2 q^2=0,$$
which is a cubic equation in $r_h^2$. The solutions are given by
\begin{eqnarray}
    r_{h,n}^2&=&2\sqrt{\frac{\frac{L^4}{3}+L^2 m-j^2 m-2 L^2 q}{3}}\cos\Bigg[\frac{1}{3}\arccos{\left(\frac{2 j^2 L^2 m-3 j^2 q^2+\frac{2 L^6}{9}+L^4 m-2 L^4 q+3 L^2 q^2}{2 j^2 m-\frac{2 L^4}{3}-2 L^2 m+4 L^2 q}\right)}\nonumber\\
    &&\sqrt{\frac{3}{\frac{L^4}{3}+L^2 m-j^2 m-2 L^2 q}}-\frac{2\pi n}{3}\Bigg],\;\;\;n=0,1,2, \label{horizon_analytical}
\end{eqnarray}
which indicates that the black hole has two horizons because of the two real positive roots. The constants $(m,q)$ are related to ($M,Q$) as given in \ref{MandQ}. As $j\to 0$, the horizon radius reduces to the five-dimensional charged AdS black hole. For both $q,j\to 0$, we have the five-dimensional Schwarzschild AdS black hole. In addition to $q,j\to 0$, if we take $L\to \infty$, we have $r_h^2=m$, which is the radius of the five-dimensional Schwazschild-Tangherlini spacetime. We plot in  \ref{fig:horizon}, the Cauchy horizon (blue dashed line) and the event horizon (black solid lines) with respect to the variation of the charge parameter (left figure) and the rotation parameter (right figure), respectively. To show the parametric bounds on $q$ and $j$, we plot the charge parameter $q$ as a function of the rotation parameter in \ref{fig:Plot1}. The region below the red line corresponds to the allowed region, whereas the region above the red line is forbidden for the parameters representing the black holes.

\section{Deflection angle in a five-dimensional charged equally rotating black hole with a cosmological constant}
In this section, we discuss the deflection angle of light rays when they pass by a black hole. We also investigate the strong field gravitational lensing phenomena in the five-dimensional, charged equally rotating CLP black hole. We see the effect of the charge $q$ and the rotation parameter $j$. For calculation simplicity, we restrict ourselves to the case of the equatorial plane, $\theta=\frac{\pi}{2}$. Therefore, ~\ref{metric} reduces to
  \begin{eqnarray}
\mathrm{ds}^2&=&\Big(-f(r)+\Sigma^2(r) \Omega^2(r)h(r)\Big)\mathrm{dt}^2+g(r)\mathrm{dr}^2+\Sigma^2(r)h(r)\mathrm{d\psi}^2+\frac{1}{4}r^2\Sigma^2(r)\mathrm{d\phi}^2\nonumber\\&&-2\Sigma^2(r)h(r)\Omega(r)\mathrm{d\psi}\mathrm{dt}.
  \label{metric'}
  \end{eqnarray}

The components of the metric tensor are written in a more compact functional form as
  \begin{eqnarray}
     g_{tt}&=&\partial_t\cdot\partial_t=-f(r)+\Sigma^2(r) \Omega^2(r)h(r)=-A(r),\nonumber \\ g_{rr}&=&\partial_r\cdot\partial_r=g(r)=B(r),\nonumber \\
     g_{\phi \phi}&=&\partial_\phi\cdot\partial_\phi=\frac{1}{4}r^2\Sigma^2(r) =C(r),\\
      g_{\psi \psi}&=&\partial_\psi\cdot\partial_\psi=\Sigma^2(r)h(r)=D(r),\nonumber \\
     g_{\psi t}&=&\partial_t\cdot\partial_\psi=-\Sigma^2(r) \Omega(r)h(r)= g_{t \psi }=-H(r). \nonumber
  \end{eqnarray}
  We can rewrite \ref{metric} as
  \begin{equation}
  \label{final_metric}
\mathrm{ds}^2=-A(r)\mathrm{dt}^2+B(r)\mathrm{dr}^2+C(r)\mathrm{d\phi}^2+D(r)\mathrm{d\psi}^2-2H(r)\mathrm{d\psi}\mathrm{dt}.
  \end{equation}
It is to be mentioned here that the coordinate $(t,r,\phi,\psi)$ are now rescaled to be $t\to t/\sqrt{M}$, $r\to r/\sqrt{M}$, $\phi\to \phi/\sqrt{M}$, and $\psi\to \psi/\sqrt{M}$. Similarly, the parameters defining the black hole are rescaled in a dimensionless way as follows: $Q\to Q/M$, $j\to j/\sqrt{M}$ and $L\to L/\sqrt{M}$. Therefore from now on, all the coordinates including the black hole parameters are understood as dimensionless quantities.
The above $ansatz$ has residual gauge freedom $\psi \to \psi+\alpha t$ and $\Omega \to \Omega+\alpha$, here $\Omega$ is the angular velocity and it is symmetric for $t \to -t$, $\phi \to \phi+2\pi$ and $\psi \to \psi+\pi$. Killing vectors are $\eta^\mu_t=\delta^\mu_t$, $\eta^\mu_\phi=\delta^\mu_\phi$ and $\eta^\mu_\psi=\delta^\mu_\psi$.
    Equations of motion are
    \begin{eqnarray}
      \mathcal{E}&=&-g_{0\mu}\Dot{x^\mu}=- g_{tt}\Dot{t}
        - g_{t \psi} \Dot{\psi}=A(r)\Dot{t}+H(r)\Dot{\psi}, \nonumber  \\
       L_{\phi}&=&g_{3\mu}\Dot{x^\mu}=g_{\phi \phi}\Dot{\phi}=C(r)\Dot{\phi}, \label{conserveq}\\
        L_{\psi}&=&g_{4\mu}\Dot{x^\mu}= g_{\psi \psi}\Dot{\psi}+g_{\psi t}\Dot{t}=D(r)\Dot{\psi}-H(r)\Dot{t}. \nonumber
    \end{eqnarray}

\begin{figure}
    \centering
    \includegraphics[scale=.6]{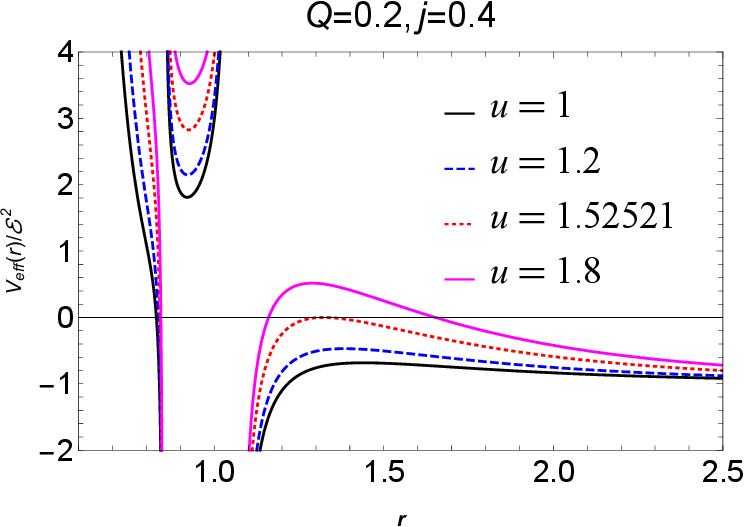}\hspace{0.1cm}
    \includegraphics[scale=.6]{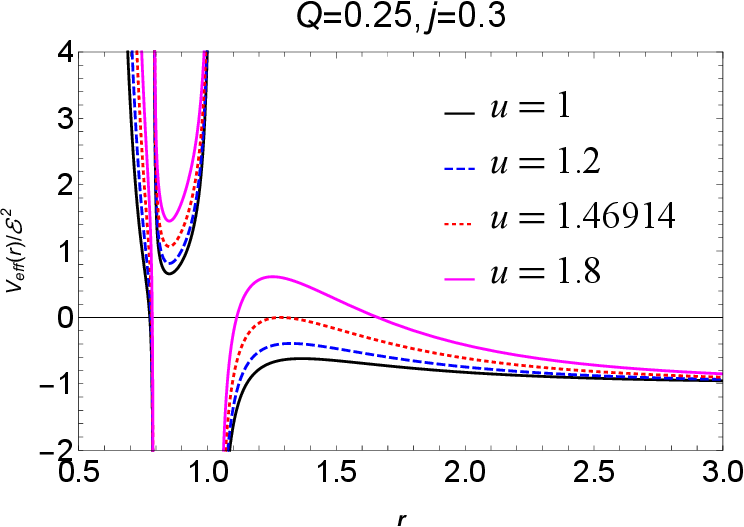}
    \caption{The behaviour of the effective potential $V_{eff}$ with respect to the dimensionless radial coordinate $r$ for different values of the impact parameter $u$.}
    \label{fig:veff}
\end{figure}

\begin{figure}[h]
    \centering
     \includegraphics[scale=.6]{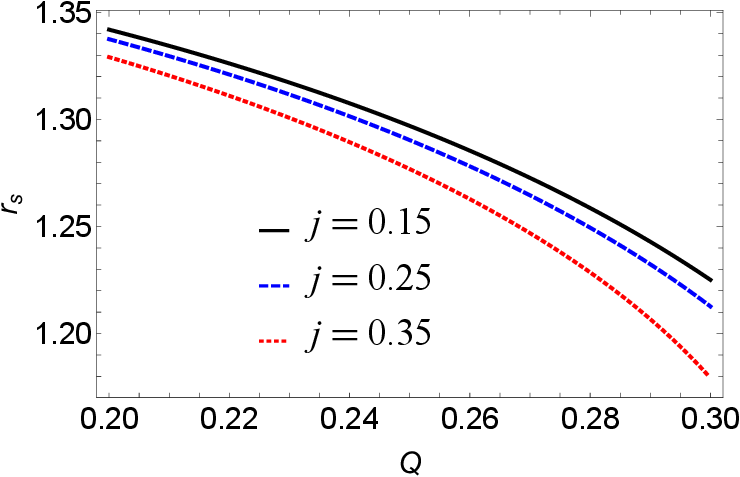}\hspace{0.1cm}
    \includegraphics[scale=.6]{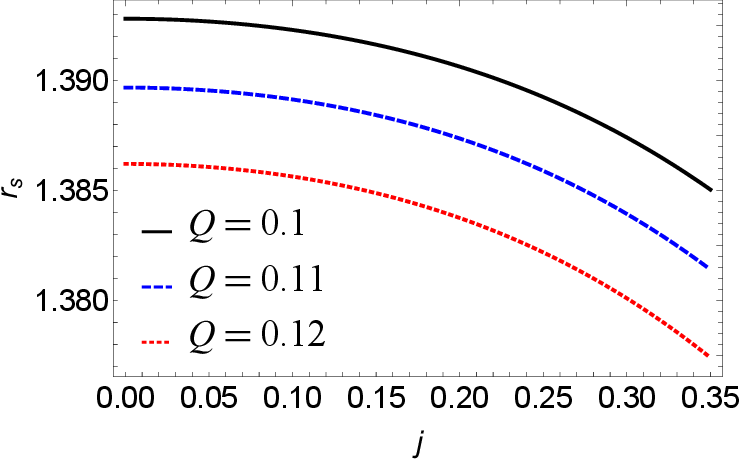}
     \caption{The variations of the photon sphere radius $r_s$ with respect to the charge parameter $Q$ (the left plot) and the rotation parameter $j$ (the right plot). }
    \label{fig:IP}
\end{figure}

\begin{figure}[h]
    \centering
    \includegraphics[scale=.6]{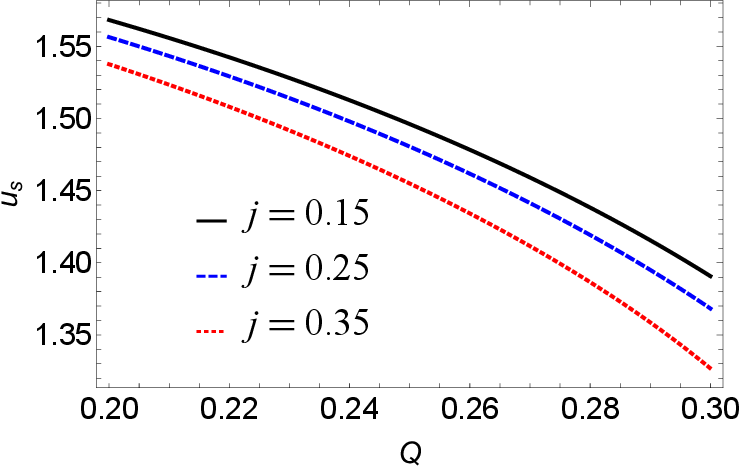}\hspace{0.5cm}
    \includegraphics[scale=.6]{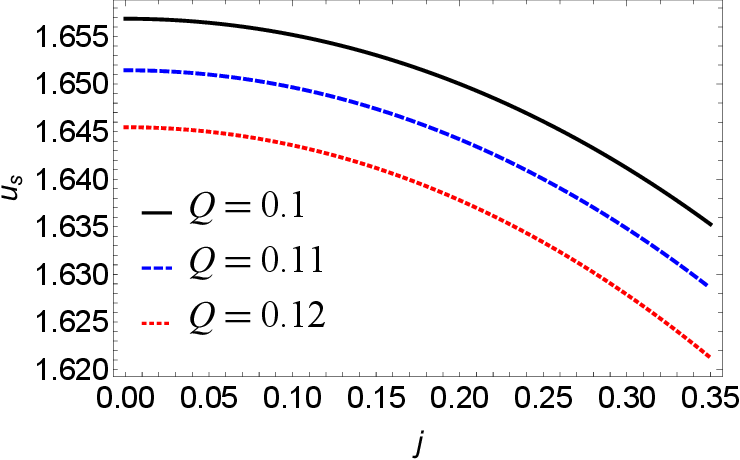}
    \caption{The variations of the impact parameter $u_s$ with respect to the charge parameter $Q$ (the left plot) and the rotation parameter $j$ (the right plot). }   
    \label{fig:IPP}
\end{figure}

Rearranging \ref{conserveq} we have
   \begin{eqnarray}
        \Dot{\psi}&=&\frac{A(r)L_\psi+H(r)\mathcal{E}}{H^2(r)+A(r)D(r)}, \nonumber \\
     \Dot{t}&=&\frac{D(r)\mathcal{E}-H(r)L_\psi}{H^2(r)+A(r)D(r)}, \label{coordeq}\\
     \Dot{\phi}&=&\frac{L_{\phi}}{C(r)} , \nonumber
  \end{eqnarray}
whereas the radial null geodesic is obtained by using $ds^2=0$
\begin{equation}
\label{rdot}
    \Dot{r}^2=\frac{1}{B(r)}\Bigg[\frac{(D(r)\mathcal{E}-2H(r)L_\psi)
    \mathcal{E}-A(r)L^{2}_{\psi}}{H^2(r)+A(r)D(r)}-\frac{L^2_\phi}{C(r)}\Bigg].
\end{equation}
The geodesic equation of motion for $\theta$ is
\begin{equation}
\label{geq}
  \frac{C^{'}(r)}{2 C(r)}  \frac{\mathrm{d\theta}}{\mathrm{d\lambda}} \frac{\mathrm{dr}}{\mathrm{d\lambda}} -\frac{H(r)\sin{\theta}}{2 C(r)}\frac{\mathrm{d\phi}}{\mathrm{d\lambda}} \frac{\mathrm{dt}}{\mathrm{d\lambda}} 
  +\frac{\cos{\theta}\sin{\theta}(-C(r)+D(r))}{C(r)}\Big(\frac{\mathrm{d\phi}}{\mathrm{d\lambda}} \Big)^2
  +\frac{D(r)\sin{\theta}}{2C(r)}\frac{\mathrm{d\phi}}{\mathrm{d\lambda}}\frac{\mathrm{d\psi}}{\mathrm{d\lambda}} =0.
\end{equation}
For $\theta=\pi/2$, \ref{geq} reduces to
\begin{equation}
 \frac{\mathrm{d\phi}}{\mathrm{d\lambda}}  \Big( -H(r)\frac{\mathrm{dt}}{\mathrm{d\lambda}}+D(r)\frac{\mathrm{d\psi}}{\mathrm{d\lambda}}\Big) =0,
\end{equation}
which gives us either $\frac{\mathrm{d\phi}}{\mathrm{d\lambda}} =0$, i.e., $L_\phi=0$ or $ L_\psi=\Big( -H(r)\frac{\mathrm{dt}}{\mathrm{d\lambda}}+D(r)\frac{\mathrm{d\psi}}{\mathrm{d\lambda}} \Big) =0$. In order to get the simpler results, we shall consider the case when $L_{\psi}=0$, which means the total angular momentum of the photon $J$ is equal to $L_{\phi}$ and the effective potential $V_{eff}=-\Dot{r}^2$, where $\Dot{r}^2$ under such conditions is given by
\begin{equation}
\label{radial}
    \Dot{r}^2=\frac{1}{B(r)}\Bigg[\frac{D(r)\mathcal{E}^2}{H^2(r)+A(r)D(r)}-\frac{L^2_\phi}{C(r)}\Bigg].
\end{equation}
To compute the impact parameter, we use $V_{eff}=0$,
\begin{equation}
    \label{impact}
    u=J=L_{\phi}/\mathcal{E}=\sqrt{\frac{C(r_{0})D(r_{0})}{H^2(r_{0})+A(r_{0})D(r_{0})}},
\end{equation}
where $u$ denotes the impact parameter. The variation of $V_{eff}$ with respect to radial distance $r$ is shown in \ref{fig:veff} for different values of the impact parameter $u$ at fixed $Q$ and fixed $j$ values. We observe that at $u_s=u_{sc}$, we have the critical impact parameter (red solid line), which depends on the parameters $Q$ and $j$. At $u_s=u_{sc}$ we have the unstable circular orbits along which the photon propagates. The photon orbit radius as a function of the charge $Q$ and the rotation parameter $j$ is depicted in \ref{fig:IP}. The photon orbit decreases quickly when plotted as a function of the charge $Q$. It decreases slowly for lower values of $j$ and for higher values of $j$, the decrease in the photon orbit radius is enhanced. Similar is the case for the impact parameter, as shown in \ref{fig:IPP}.

%The radial geodesic equation is
%\begin{equation}
   % \frac{d^2r}{d \lambda^2}+\frac{A^{\prime}(r)}{2 B(r)}\Big(\frac{dt}{d\lambda}\Big)^2+\frac{B^{\prime}(r)}{2 B(r)}\Big(\frac{dr}{d\lambda}\Big)^2-\frac{C^{\prime}(r)^2}{2B(r)C(r)^2}J^2=0
%\end{equation}
The equation of the circular orbit of a photon is obtained using $V^{\prime}_{\text{eff}}=0$,
\begin{equation}
\label{COE}
    {C}(r_0) \left[{D}(r_0)^2 A'(r_0)-H(r_0)^2 {D}'(r_0)+2 {D}(r_0) H(r_0) H'(r_0)\right]-{D}(r_0) {C}'(r_0) \left[A(r_0) {D}(r_0)+H(r_0)^2\right]=0.
\end{equation}
where $r_0$ is the closest of the photon spheres. Eqs.~(\textcolor{blue}{26}) and (\textcolor{blue}{27}) are more complex than those in the usual spherical, symmetric black hole spacetime. As a limiting case, when the parameter $j \to 0$, we have $H(r)\to 0$ and therefore, \ref{COE} has a much simpler form,
\begin{equation}
\label{COE-lim}
    {C}(r_0)A'(r_0)-{C}'(r_0) A(r_0)=L^2 \left(4 q^2+6 q x+x (2 x-3)\right)+x^3=0,\;\;x=r_0^2.
\end{equation}
The above equation is cubic in nature and the roots are easily derivable as
\begin{eqnarray}
    \label{solCOE_j0}
 x_k=2\sqrt{-\frac{\mathcal{P}}{3}}\cos\left[\frac{1}{3}
  \arccos{\left(\frac{3\mathcal{Q}}{2\mathcal{P}}\sqrt{\frac{-3}{\mathcal{P}}}\right)}-\frac{2\pi k}{3}\right],\;\;\;k=0,1,2,
\end{eqnarray}
where $\mathcal{P}=\left(6q-3\right)L^2-\frac{4L^4}{3}$, $\mathcal{Q}=4l^2q^2+2L^4\left(1-2q\right)-\frac{16L^6}{27}$. Obviously, for $q=0$, we have only one root that corresponds to $x=\sqrt{L^2 \left(L^2+3\right)}-L^2$. However, we cannot determine the photon sphere radius analytically for $j\neq 0$ so we have computed it numerically. The value of the impact parameter for $j=0$ is computed to be
\begin{eqnarray}
    \label{impact_2}
    u_s|_{j=0}&=&\frac{L \mathcal{X}_3^2}{2 \sqrt{3} {\mathcal{X}_2}^{1/3} \left(27 L^2 q^2 \mathcal{X}_2+L^2 \mathcal{X}_3^2 \left(4 L^2-18 q+{\mathcal{X}_2}^{1/3}+9\right)+\mathcal{X}_3 \left(4 L^4 \mathcal{X}_2^{2/3}-2 L^2 \mathcal{X}_2+\mathcal{X}_2^{4/3}\right)\right)}, \nonumber\\
\mathcal{X}_1&=&\sqrt{-3 L^8 \left(4 q^2-36 q+9\right)-81 L^6
     \left(8 q^2-6 q+1\right)+324 L^4 q^4}, \nonumber\\
\mathcal{X}_2&=&-8 L^6+27 L^4 (2 q-1)-54 L^2 q^2+3 \mathcal{X}_1, \nonumber\\
\mathcal{X}_3&=&4 L^4+L^2 \left(-18 q-2 {\mathcal{X}_2}^{1/3}+9\right)+\mathcal{X}_2^{2/3}.
\end{eqnarray}
\begin{figure}[h]
    \centering
     \includegraphics[scale=.62]{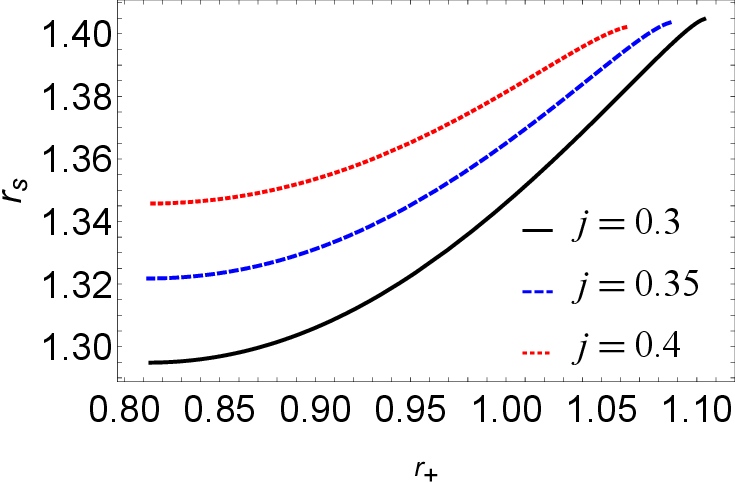}\hspace{0.1cm}
    \includegraphics[scale=.62]{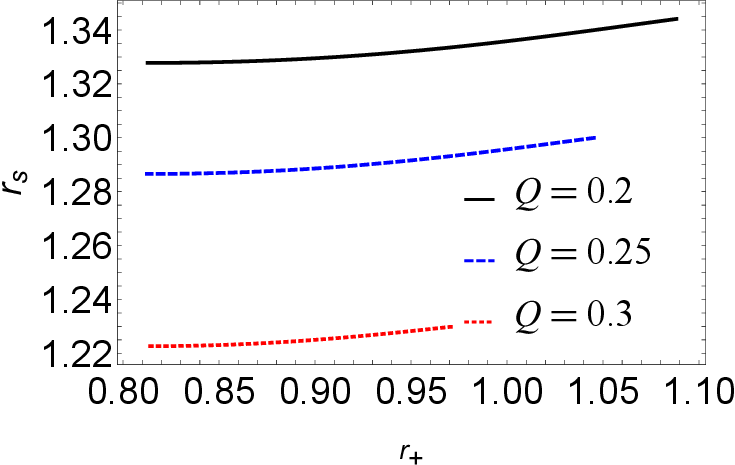}
     \caption{The behaviour of the photon orbit $r_s$ (in units of $\sqrt{M}$) with respect to the horizon radius $r_+$ (in units of $\sqrt{M}$) for different values of the charge parameter with the fixed rotation parameter $j=0.1$ (the left plot) and different values of the rotation parameter with the fixed charge parameter $Q=0.1$ (the right plot).}
    \label{fig:rsrh}
\end{figure}

\begin{figure}[h]
    \centering
     \includegraphics[scale=.62]{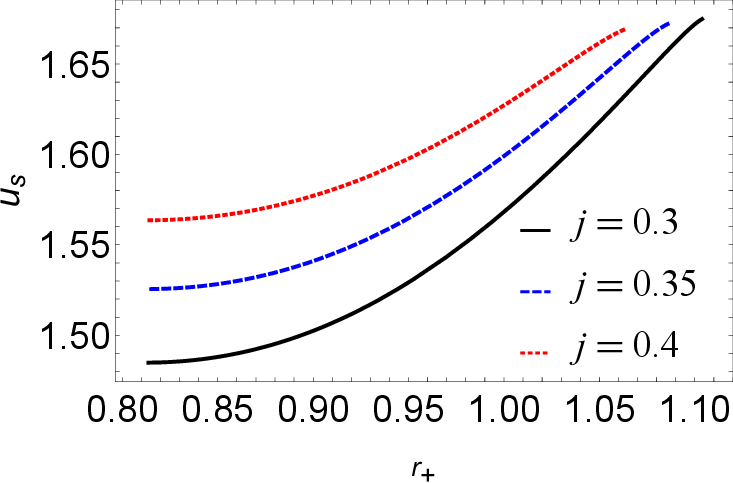}\hspace{0.1cm}
    \includegraphics[scale=.62]{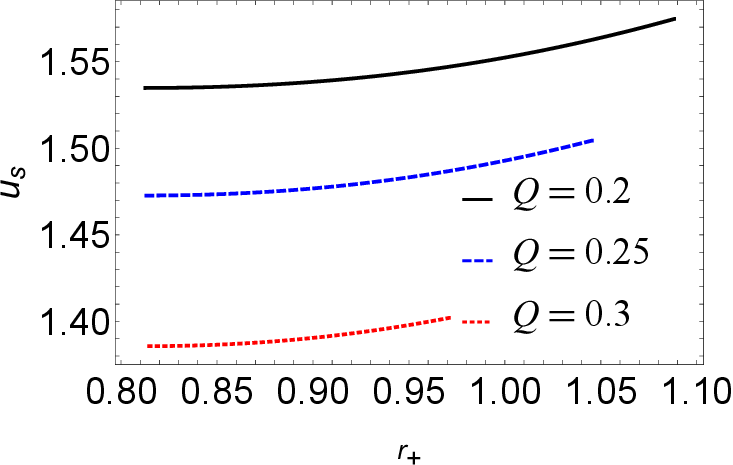}
     \caption{The behaviour of the impact parameter $u_s$ (in units of $\sqrt{M}$) with respect to the horizon radius $r_+$ (in units of $\sqrt{M}$) for different values of the charge parameter with the fixed rotation parameter $j=0.1$ (left) and for different values of the rotation parameter with the fixed charge parameter $Q=0.1$ (right). }
    \label{fig:usrh}
\end{figure}

In \ref{fig:rsrh} and \ref{fig:usrh}, the variations of the photon orbits and the impact parameter as functions of the horizon radius for a fixed $Q$ (left) and a fixed $j$ value (right) are shown, respectively. For the fixed $Q$ value, the variation of the photon orbits as well as the impact parameter is almost constant, but the gap of the plots is highly sensitive to the $j$ value. For a fixed $j$ value, the variations of both $r_s$ and $u_s$ are monotonically varying functions of the horizon radius. They increase as the horizon's radius increases. For vanishing angular momentum, i.e., for $j=0$, we have the exact expression for the photon orbits, \ref{solCOE_j0} and the minimum impact parameter, \ref{impact_2}. We observe that, as in the four-dimensional Kerr black hole spacetime, the value of the unstable circular orbit is dependent on the photon winding number in the direction of the black hole rotation $(j>0)$ or in the opposite direction of the black hole rotation (i.e., $j<0$). It is also emphasized that for the Kaluza-Klein rotating black hole with squashed horizons, the unstable circular orbits are independent of the black hole rotation, as the rotation parameter appears to be an even power of $j^2$.

Since $\Dot{r}^2$ vanishes at the closest distance $r=r_0$, from the trajectory equation we have
\begin{equation}
\label{trajectory}
    A(r_0)\Dot{t}_{0}^2+2H(r_0)^2\Dot{t}_{0}\Dot{\psi}_{0}=C(r_0)\Dot{\phi}_{0}^2+D(r_0)\Dot{\psi}_{0}^2,
\end{equation}
%Using
%\begin{equation}
 %   D(r)\Dot{\psi}=H(r)\Dot{t}
%\end{equation}
%at $r=r_0$ in \ref{trajectory}. We have
%\begin{equation}
 %   \Big(A(r_0)D(r_0)+H(r_0)^2(2H(r_0)-1)\Big)\Dot{t_0}^2=D(r_0)C(r_0)\Dot{\phi_0}^2
%\end{equation}
Using \ref{rdot}, we have
\begin{equation}
\label{R}
    \Dot{r}=\frac{\mathrm{dr}}{\mathrm{d\lambda}} =\frac{1}{\sqrt{B(r)F(r_0)F(r)}}\sqrt{F(r_0)-\frac{C(r_0)F(r)}{C(r)}}.
\end{equation}
The radial motion of the photon is governed by
\begin{equation}
 \frac{\mathrm{d\phi}}{\mathrm{dr}} = \frac{\sqrt{B(r)F(r_0)F(r)}}{C(r)}\frac{1}{\sqrt{F(r_0)-\frac{C(r_0)F(r)}{C(r)}}}
\end{equation}
and
\begin{equation}
  \frac{\mathrm{d\psi}}{\mathrm{dr}} =\frac{H(r)}{D(r)}\sqrt{\frac{B(r)F(r_0)}{F(r)}} \frac{1}{\sqrt{F(r_0)-\frac{C(r_0)F(r)}{C(r)}}}
\end{equation}
with
\begin{equation}
    \label{F}
    F(r)=\frac{H^2(r)+A(r)D(r)}{D(r)}.
\end{equation}
The deflection angles $\psi$ and $\phi$ for the photon coming from the infinite can be expressed as \cite{Bozza:2002zj}
\begin{eqnarray}
\label{integrations}
I_\psi(r_0)&=&2 \int^\infty_{r_{0}} \frac{\mathrm{d\psi}}{\mathrm{dr}}\mathrm{dr}=2\int^\infty_{r_{0}}\mathrm{dr}\frac{H(r)}{D(r)}\sqrt{\frac{B(r)F(r_0)}{F(r)}} \frac{1}{\sqrt{F(r_0)-\frac{C(r_0)F(r)}{C(r)}}},\\
I_\phi(r_0)&=&2 \int^\infty_{r_{0}} \frac{\mathrm{d\phi}}{\mathrm{dr}}\mathrm{dr}=2\int^\infty_{r_{0}}\mathrm{dr}
\frac{\sqrt{B(r)F(r_0)F(r)}}{C(r)}\frac{1}{\sqrt{F(r_0)-\frac{C(r_0)F(r)}{C(r)}}}.
\end{eqnarray}

Except for the metric function $H(r)$ which is an odd function of the rotation parameter, the other functions contain an even power of the rotation parameter. Therefore, the deflection angle corresponding to the $\phi$-angle is dependent on whether the black hole is in retrograde or in prograde motion. However, the deflection angle related to the angle $\psi$ is an even function of the rotation parameter $j$, which means that it does not matter whether the photon is winding in the direction of the black hole rotation or in the converse direction, the deflection angle is always independent of the black hole rotation. Furthermore, unlike the four-dimensional case, in the equatorial plane $(\theta=\pi/2)$, the rotation of the black hole is in the $\psi$ direction rather than in the $\phi$ direction, which is evident from the expression of the spacetime as is given in \ref{final_metric}, where the only cross-term is $dt d\psi$. This also clarifies the fact that the gravitational lensing of the charged rotating black hole in five dimensions is not the same as the strong lensing phenomena of the usual four-dimensional Kerr or the Kerr-Newman black hole. \\

On defining $z=1-\frac{r_0}{r}$, we have
\begin{equation}
\label{Iphi}
  I_\phi(r_0)=\int_0^1  R(z,r_0) f(z,r_0)\mathrm{dz}
\end{equation}
with
\begin{eqnarray}
   R(z,r_0)&=&2 \frac{r^2}{r_0 C(r)}\sqrt{B(r)F(r)C(r_0)}=P_{1}\left(r, r_{0}\right)
\label{Rz} , \\
    f(z,r_0)&=&\frac{1}{\sqrt{F(r_0)-F(r)C(r_0)/C(r)}}=P_{2}\left(r, r_{0}\right).
    \label{fz}
\end{eqnarray}
The function $R(z,r_0)$ is regular for all values of $z$ and $r_0$. From \ref{fz}, we find that  $f(z,r_0)$ diverges as $z$ tends to zero, i.e., as the photon approaches the marginally circular photon orbit. Therefore, we can split the integral \ref{Iphi} into divergent part $I_D(r_0)$ and the regular one $I_R(r_0)$
\begin{eqnarray}
\label{IDR}
I_{D,\phi}(r_0) &=& \int^1_0 R(0,r_s)f_0(z,r_0)\mathrm{dz},\\
I_{R,\phi}(r_0) &=& \int^1_0[R(z,r_0)f(z,r_0)-R(0,r_s)f_0(z,r_0)]\mathrm{dz}.
\end{eqnarray}
 We can expand the argument of the square root in $f(z,r_0)$ to the second order in $z$
\begin{equation}
    \label{fexpn}
    f_0(z,r_0)=\frac{1}{\sqrt{p(r_0)z+g(r_0)z^2}}
\end{equation}
with
\begin{equation}
    \begin{split}
    \label{coefff}
        p(r_0)&=\frac{r_0}{C(r_0)}\Big[C^\prime(r_0)F(r_0)-C(r_0)F^\prime(r_0)\Big],\\
        g(r_0)&=\frac{r_0^2}{2C(r_0)}\Big[2C^\prime(r_0)C(r_0)F^\prime(r_0)
        -2C^\prime(r_0)^2F(r_0)+F(r_0)C(r_0)C^{\prime\prime}(r_0)-C^2(r_0)F^{\prime\prime}(r_0)\Big].
    \end{split}
\end{equation}
To obtain \ref{fexpn} and \ref{coefff} we expanded $f(z,r_0)$ around $z=0$ upto order $z^2$. If we use \ref{COE}, one obtains $F(r)C^\prime(r)=C(r)F^\prime(r)$ which leads to $p(r_0)=0$ and $r_s$ is the largest root of $p(r_0)$. This means that the leading term of the divergence in $f_0(z,r_0)$ is $z^{-1}$ and the integral \ref{Iphi} diverges logarithmically. Thus, in the strong field region, the deflection angle in the direction can be approximated very well as \cite{Bozza:2002zj}:
\begin{equation}
    \label{deflection}
    \alpha_{D,\phi} (\theta)=-\Bar{a}\ln\Big(\frac{\theta D_{OL}}{u_{s}}-1\Big)+\Bar{b}+O(u-u_s),
\end{equation}
where
\begin{equation}
\label{ab}
    \begin{split}
   \Bar{a}&=\frac{R(0,r_s)}{2\sqrt{g(r_s)}},\\
   \Bar{b}&=-\pi+b_R+\Bar{a}
       \ln\left(\frac{r^2_s[C^{\prime\prime}(r_s)F(r_s)-C(r_s)F^{\prime\prime}(r_s)]}
                {u_{s}F(r_s)\sqrt{F(r_s)C(r_s)}}\right), \\
   b_R&=I_R(r_s),\quad\quad u_s=\sqrt{\frac{C(r_s)}{F(r_s)}}.
    \end{split}
\end{equation}
Here the quantity $D_{\text{OL}}$ is the distance between observer and gravitational lens, $\theta=u/D_{\text{OL}}$
is the angular separation between the lens and the image. The subscript ``$s$" represents the
evaluation at $r=r_s$.
Likewise, we can compute the corresponding  $\psi$ following \ref{Iphi}. We can write
\begin{equation}
    I_{\psi}(r_0)=\int_0^1 S(z,r_0)f(z,r_0)\mathrm{dz}
\end{equation}
with $f(z,r_0)$ given by \ref{fz} and $S(z,r_0)$ as
\begin{equation}
    \label{Sz}
    S(z,r_0)=2 \frac{r^2 H(r)}{r_0 D(r)}\sqrt{\frac{B(r)F(r_0)}{F(r)}}.
\end{equation}
The function $S(z,r_0)$ is regular for all values of $z$ and $r_0$. Whereas $f(z,r_0)$ is divergent as for the case of  $I_{D,\phi}(r_0)$, and we can follow the same expansion as well but replace $\Bar{a}$ with $\Bar{a}_\psi$,
\begin{eqnarray}
  & \Bar{a}_\psi=\frac{S(0,r_s)}{2\sqrt{g(r_s)}}, \label{abarpsi}\\
 &I_{D,\psi}(r_0)=\int^1_0 S(0,r_s)f_0(z,r_0)\mathrm{dz} ,  \label{IDR'}\\
 & I_{S,\psi}(r_0)=\int^1_0[S(z,r_0)f(z,r_0)-R(0,r_s)f_0(z,r_0)]\mathrm{dz}.
\end{eqnarray}
\begin{figure}[h]
    \centering
    \includegraphics[scale=.6]{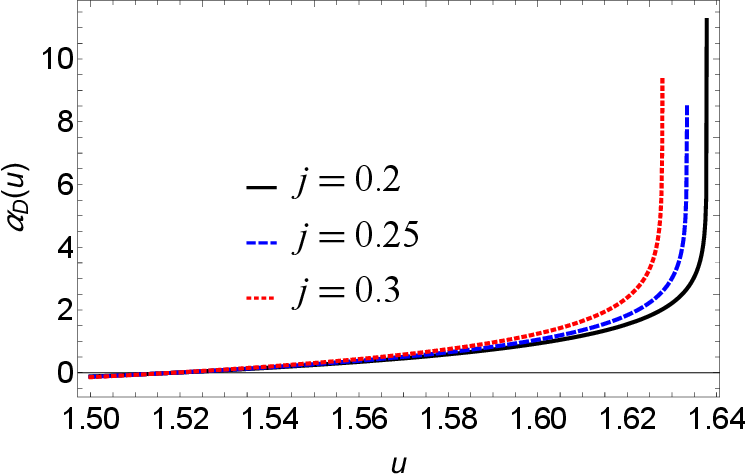}\hspace{0.1cm}
    \includegraphics[scale=.58]{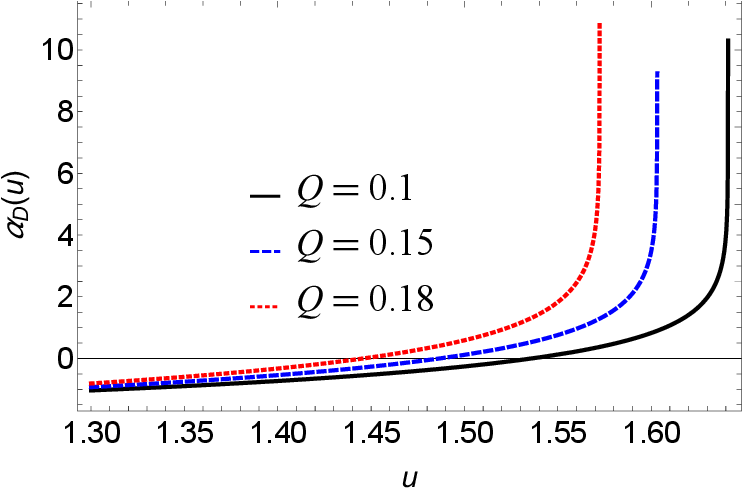}
    \caption{ The variation of the deflection angle $\alpha_D (u)$ with respect to the impact parameter $u$. We have set $Q=0.12$ in the left plot and $j=0.3$ in the right one.
    }
    \label{fig:deflection angle}
\end{figure}
\begin{figure}[h]
    \centering
    \includegraphics[scale=.6]{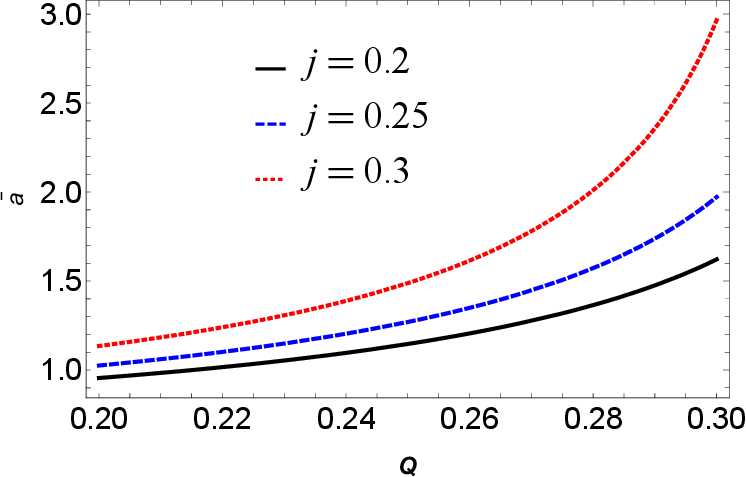}\hspace{0.1cm}
    \includegraphics[scale=.58]{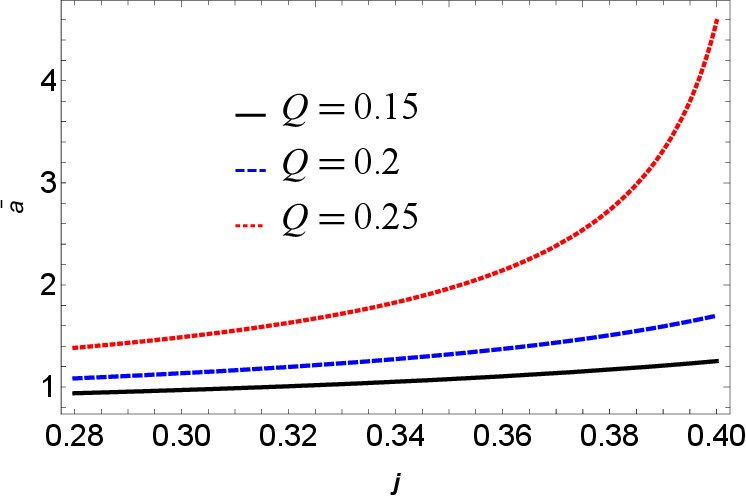}
    \caption{The variations of the deflection coefficient $\Bar{a}$ with respect to the charge parameter $Q$ (the left plot) and the rotation parameter $j$ (the right plot). }
    \label{fig:a_bar}
\end{figure}
\begin{figure}[h]
    \centering
    \includegraphics[scale=.61]{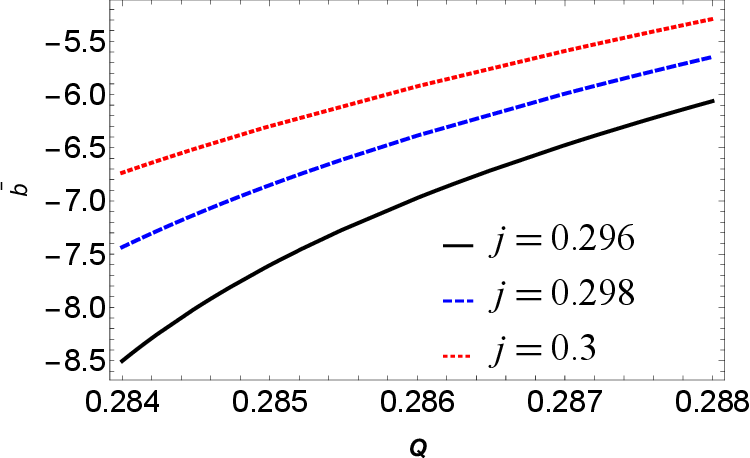}\hspace{0.1cm}
    \includegraphics[scale=.61]{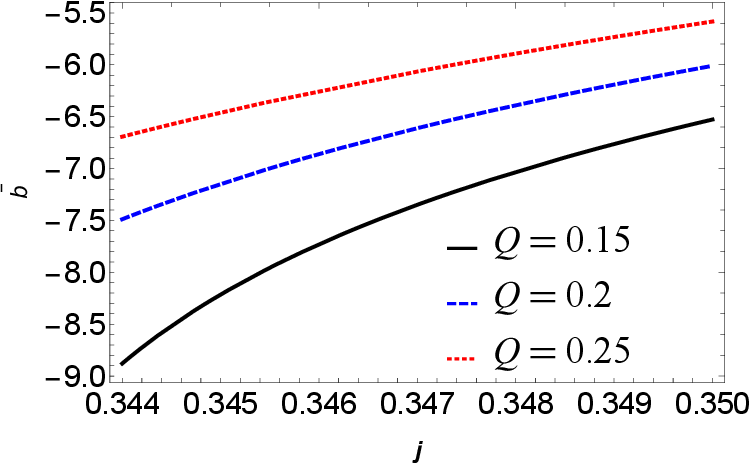}
    \caption{The variations of the deflection coefficient $\Bar{b}$ with respect to the charge parameter $Q$ (the left plot) and the rotation parameter $j$ (the right plot).}
    \label{fig:b_bar}
\end{figure}
The study of the deflection angle only in the $\phi$ direction would be focused on our recent purpose since it helps in observing real astronomical observations. One can likewise determine the integration regarding the strong lensing deflection angle in the $\psi$ direction, which we denote by $I_{\psi(r_0)}$. This term also has the coefficients, which likely have the form of $\Bar{a}$ and $\Bar{b}$ but differ slightly. We should mention that the deflection angle in the $\phi$-driection also has a term that diverges logarithmically. Nevertheless, it should not be used in determining the strong deflection phenomena from observational perspectives. In \ref{fig:deflection angle}, we plot the deflection angle with respect to the impact parameter for different values of the rotation parameter $j$ (the left figure) and the charge parameter $Q$ (the right figure). We can see from the figure that the deflection angle increases monotonically with the critical impact parameter $u$ and diverges for large $u$. Similarly, the variations of the coefficients $\Bar{a}$ and $\Bar{b}$ with respect to the rotation parameter and the charge parameter are shown in \ref{fig:a_bar} and \ref{fig:b_bar}, respectively. We observe a monotonic variation for the coefficient $\Bar{a}$ it increases with increasing $Q$ and $j$, whereas the coefficient $\Bar{b}$ increases with $Q$ but decreases with $j$.

\section{Observables and relativistic images}
\label{Sec4}
The description of the gravitational lensing phenomena is determined through the lens equation. Many methods for calculating the lens equation depend on the choice of parameters we are interested in. In our calculation, we choose the gravitational lens, where the black hole is placed in the middle of the configuration so that the observer is situated at one side, and the illuminating source of light is at another side. The light rays emanate from the source (S) and their actual paths are deviated due to the presence of the central black hole with high curvature. Then after making one or more windings, the light rays finally reach the observer position (O). In the picture of the black hole lens, the optical axis OL is the line connecting the black hole, the observer and the image, which will deviate at an angle $\theta$ with respect to OL. On the other hand, the illuminating source of light is oriented at an angle $\beta$ with respect to OL. The light emanating by the source is detected by the observer by an angle $\alpha_{D}(\theta)$.

\begin{figure}[h]
    \centering
    \includegraphics[scale=.6]{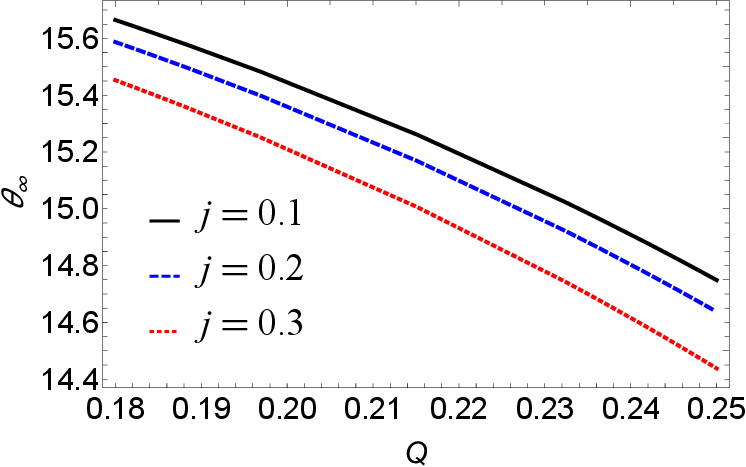}\hspace{0.1cm}
    \includegraphics[scale=.6]{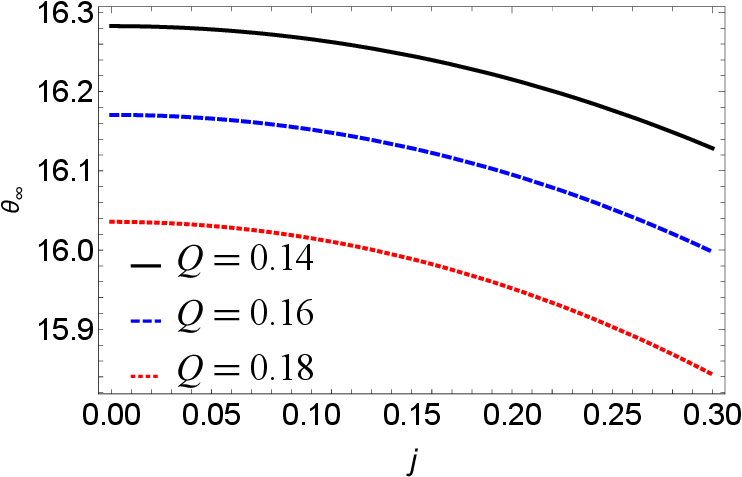}\\
     \includegraphics[scale=.6]{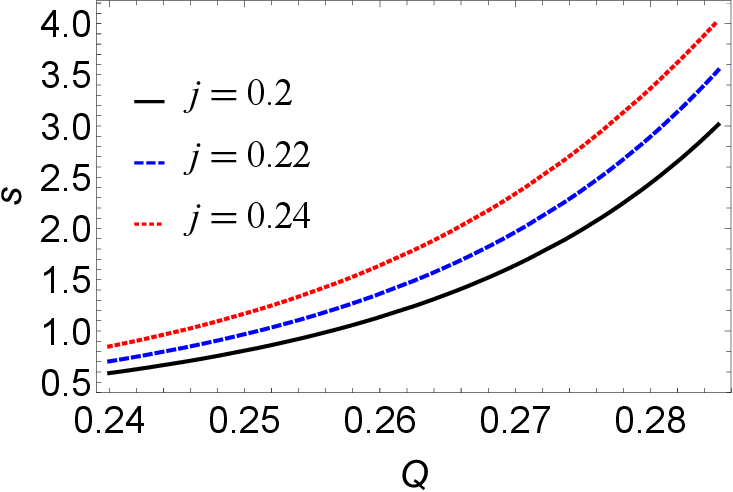}\hspace{0.1cm}
    \includegraphics[scale=.6]{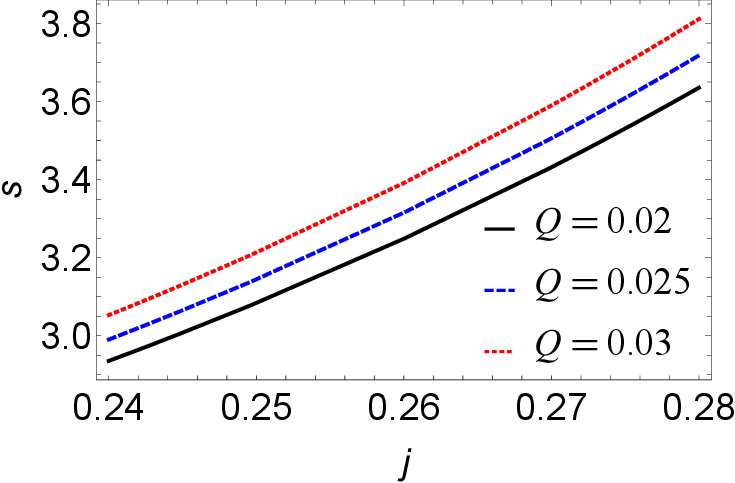}\\
     \includegraphics[scale=.6]{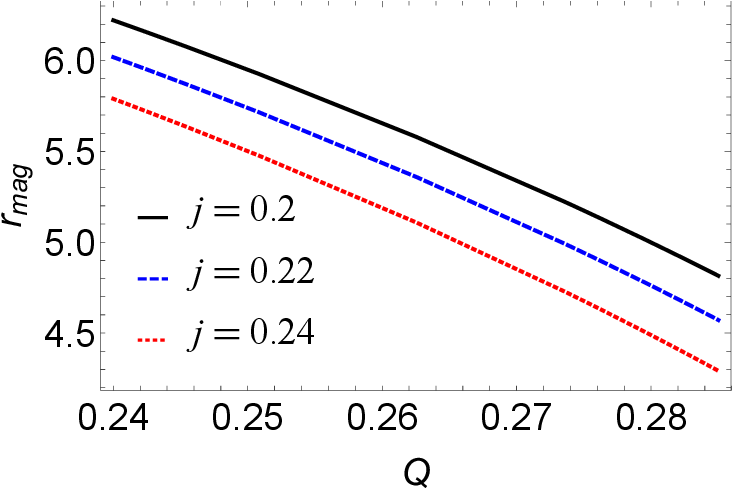}\hspace{0.1cm}
    \includegraphics[scale=.62]{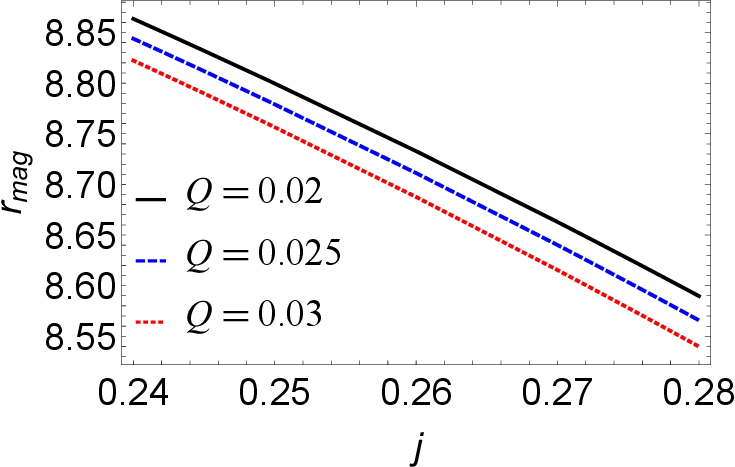}
    \caption{ Plot of the lensing variables for Sgr A* with respect to the charge parameter $Q$  (the left plot) and the rotation parameter $j$ (the right plot).}
    \label{fig:deflection_m87}
\end{figure}

\begin{figure}[h]
    \centering
    \includegraphics[scale=.62]{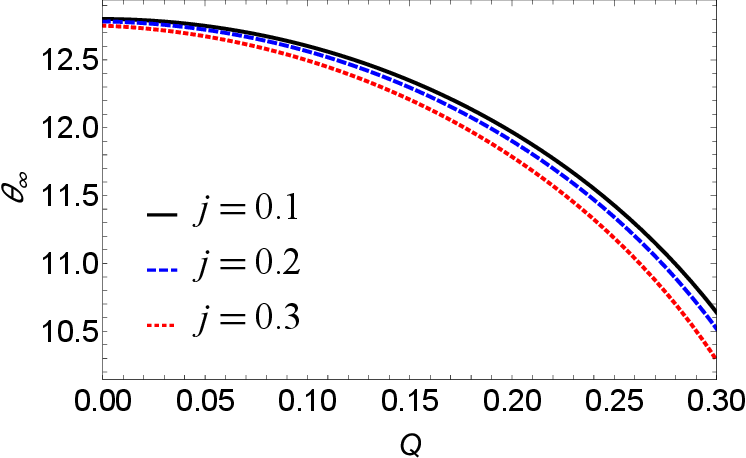}\hspace{0.1cm}
    \includegraphics[scale=.62]{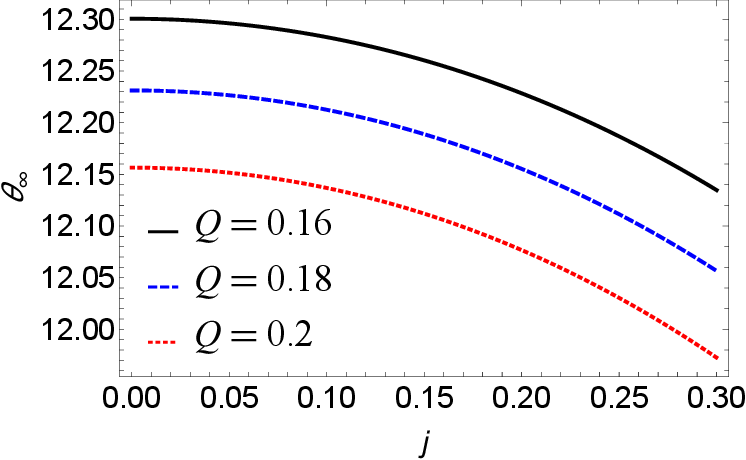}\\
    \includegraphics[scale=.6]{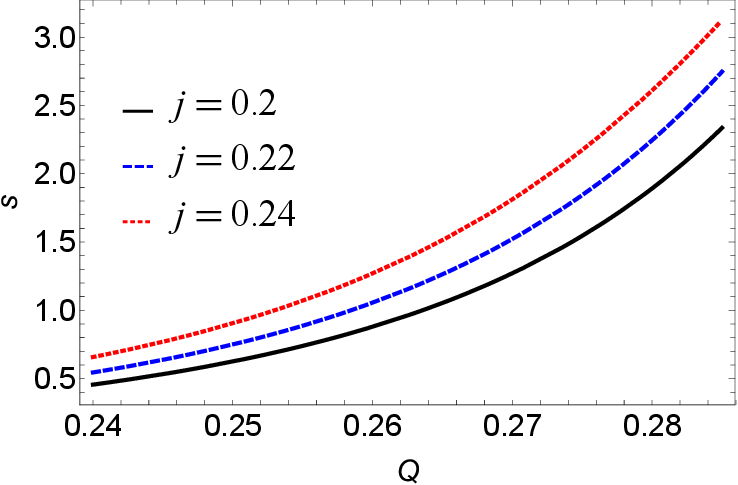}\hspace{0.1cm}
    \includegraphics[scale=.6]{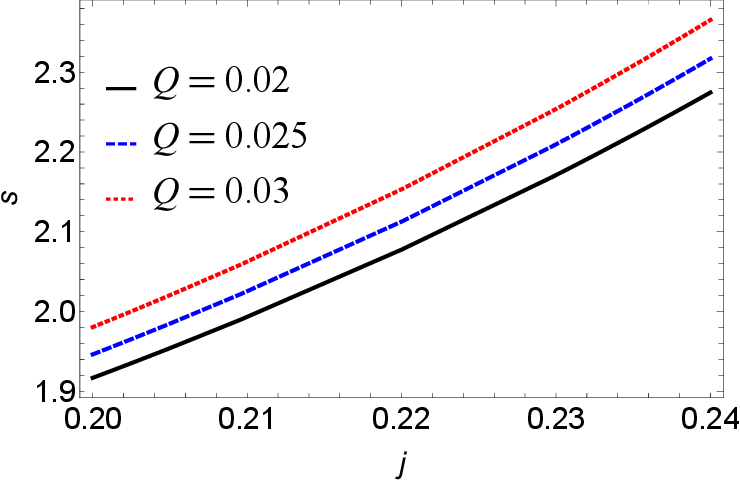}\\
    \includegraphics[scale=.6]{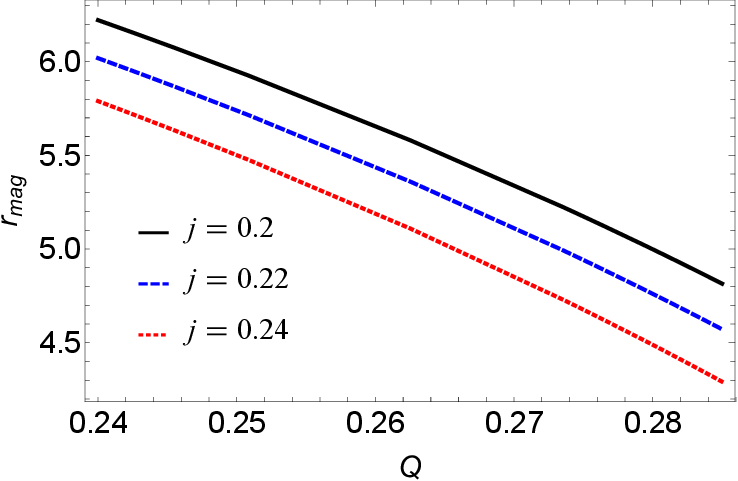}\hspace{0.1cm}
    \includegraphics[scale=.61]{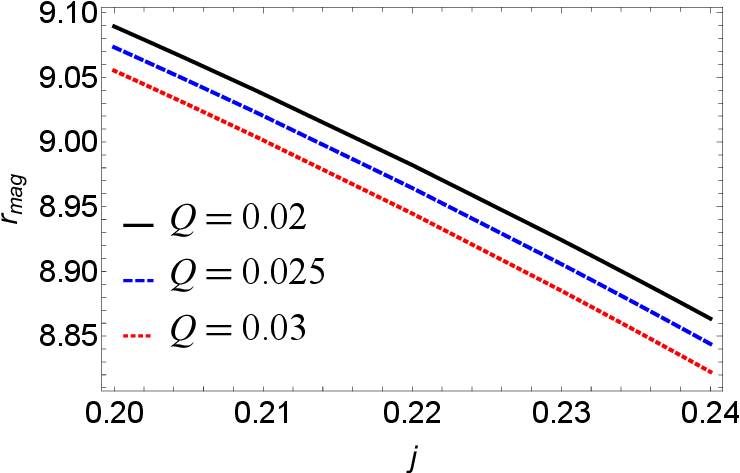}
    \caption{ Plot of the lensing variables for M87 with respect to the charge parameter $Q$  (the left plot) and the rotation parameter $j$ (the right plot).}
    \label{fig:magnification_m87}
\end{figure}

Among various mathematical formulations, the method given by Ohanian was mostly used in the literature as the lens equation \citep{Ohanian:1987pc} to approximate the position of the source, lens and observer:
\begin{eqnarray}
\label{oho}
\xi &=& \frac{D_{OL}+D_{LS}}{D_{LS}}\theta-\alpha_{D}(\theta).
\end{eqnarray}
The angle $\xi\in[-\pi,\pi]$ is used for the optical axis and the source orientations, whereas $D_{OL}$ is the distance connecting the source and the observer, and $D_{LS}$ is the line between the source and lens. The angles $\xi$ and $\beta$ are found to follow the relation \cite{Ohanian:1987pc, Ghosh:2020spb}
\begin{eqnarray}\label{rel}
\frac{D_{OL}}{\sin(\xi-\beta)} &=& \frac{D_{LS}}{\sin \beta}.
\end{eqnarray}
For the completeness of the calculations and to have physical realization, the angles $\theta$, $\xi$, and $\beta$ are considered to be smaller because the formation of the relativistic images are dominant feature of the strong lensing phenomena. The light emanating from the source encounters the black hole and makes many winding or loops before it leaves such that the deflection angle $\alpha $ is framed as $2n\pi + \Delta\alpha _n$, where $n \in N $ is an integer and is accounted for the loop counting and $ 0<\Delta\alpha _n \ll 1$. Therefore, following \ref{oho} along with \ref{rel}, for the smaller values of $\theta$, we have the lens equation
\begin{eqnarray}
\label{lensequation}
\beta &=& \theta -\frac{D_{LS}}{D_{OL}+D_{LS}} \Delta\alpha _n.
\end{eqnarray}
We can use \ref{lensequation} to know information about the image formation. We know that when the impact parameter $u\to$ $u_s$, the deflection angle $\alpha_D(\theta)$ becomes divergent. Therefore, for each loop formed during the winding near the event horizon of the black hole, we certainly have one $u_s$ at which the light rays reach from the lens to the observer. Accordingly, we have infinitely many images that are formed on both sides of the black hole. Now \ref{deflection} together with $\alpha_{D}(\theta_n{^0}) = 2n\pi $ is rewritten as
\begin{eqnarray}\label{theta}
\theta_n{^0} &=& \frac{u_s}{D_{OL}}(1+\rm e_n),
\end{eqnarray}
where
\begin{eqnarray}
\rm e_n &=& \rm e^{\frac{\Bar{b}-2n\pi}{\Bar{a}}}.
\end{eqnarray}
%%%%%%%%%%%%%%%%%%%%%%%%%%%%%%%%%%%%%%%%%%%%%%%%%%%%%%%%%%%%%%%%%%%%%
Now a Taylor series expansion of $\alpha_{D}(\theta)$ around $\theta_n{^0}$ up to first order in $(\theta-\theta_n{^0})$ may be approximated as \cite{Ghosh:2020spb}
\begin{eqnarray}
\alpha_{D}(\theta) &=& \alpha_{D}(\theta_n {^0}) +\frac{\partial \alpha_{D}(\theta)}{\partial \theta } \Bigg |_{\theta_n{^0}}(\theta-\theta_n{^0})+\mathcal{O}(\theta-\theta_n{^0})^2.
\end{eqnarray}
Using \ref{theta} along with the condition $\Delta\theta_n= \theta-\theta_n{^0} $, we have the difference of the deflection angle
\begin{eqnarray}
\Delta\alpha_n &=& -\frac{\Bar{a}D_{OL}}{u_s \rm e_n}\Delta\theta_n.
\end{eqnarray}
Therefore, the final lens equation \ref{lensequation} is approximated to be \cite{Ghosh:2020spb}
\begin{eqnarray}
\label{final}
\beta &=&  \theta_n{^0} + \Delta\theta_n +\frac{D_{LS}}{D_{OL}+D_{LS}}\Bigg(\frac{\Bar{a} D_{OL}}{u_s \rm e_n}\Delta\theta_n\Bigg).
\end{eqnarray}
We ignore the second term in \ref{final} since its contribution is very less in comparison to other terms, and then by substituting $\Delta\theta_n= (\theta-\theta_n{^0}) $, we have the $\theta$-equation for the $n$-th image as
\begin{eqnarray}\label{angpos}
\theta_n &=& \theta_n{^0} + \frac{D_{OL}+D_{LS}}{D_{LS}}\frac{u_s \rm e_n}{\Bar{a}D_{OL}}(\beta-\theta_n{^0}).
\end{eqnarray}
\begin{figure}
    \centering
    \includegraphics[scale=.65]{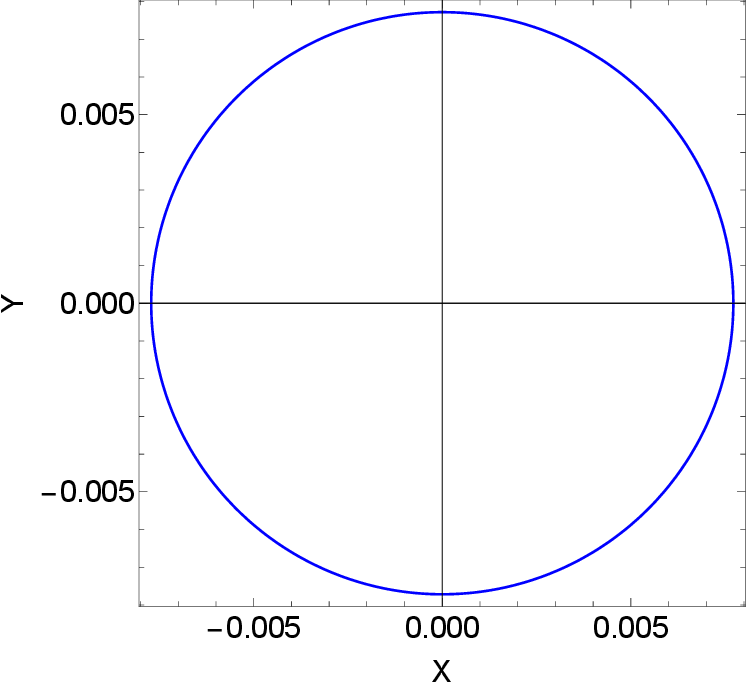}\hspace{0.1cm}
    \includegraphics[scale=.61]{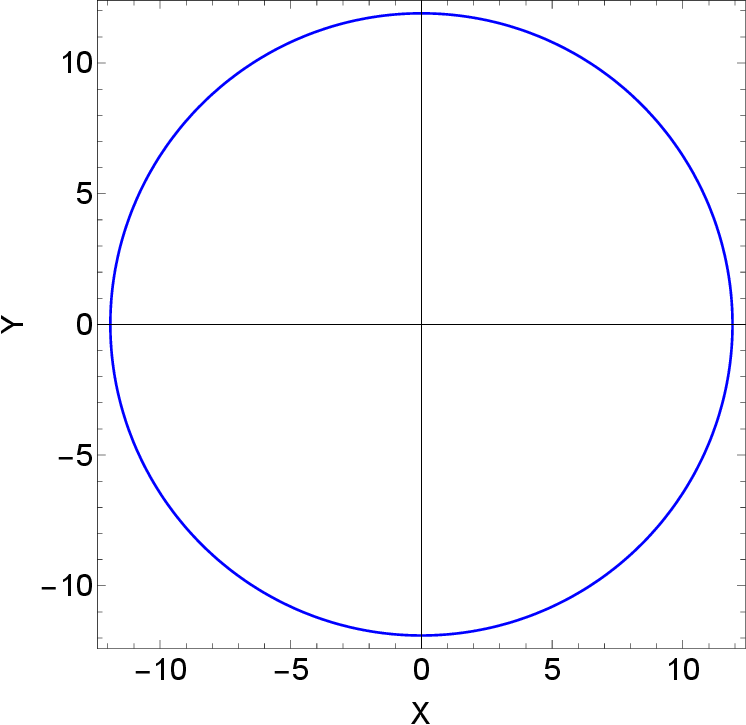}
    \caption{Plots of the outermost Einstein rings for the black holes at the centre of nearby galaxies in the framework of the Schwarzschild geometry. The left plot corresponds to the $SgrA^*$ and the right one to that of $M87$ \cite{Ghosh:2020spb}.}
    \label{ERfig}
\end{figure}
Next, we discuss the creation and subsequent formation of Einstein's ring. Among many exotic features of the lensing phenomenon, the formation of Einstein's ring is the most striking one. When the light source, lens, and observer are aligned in a most suitable way so that the whole lens system plays a role in the formation of complex multiple images \cite{Ohanian:1987pc,Virbhadra:1999nm,Virbhadra:2002ju, Virbhadra:2008ws}. We also have double Einstein's ring \cite{Gavazzi:2008} formed depending on the source we have. For relativistic Einstein's ring formation, the deflection angle $\alpha$ must have a value greater than $2\pi$. For perfect alignment of the lens and the observer, we must set $\beta=0$, so that the lens is situated at the middle of the line connecting the source and the observer. Therefore, \ref{angpos} reduces to \cite{Ghosh:2020spb}
	\begin{eqnarray}\label{Ering2}
		\theta_n^{E} &=& \left(1-\frac{2 u_s \rm e_n}{D_{OL}\Bar{a}} \right) . \left(\frac{u_s}{D_{OL}}(1+\rm e_n)\right).
	\end{eqnarray}
As a limiting case when $D_{OL} \gg u_s$, the angular radius \ref{Ering2} for Einstein's ring reduces to
	\begin{eqnarray}\label{Ering3}
		\theta_n^{E} = \frac{u_s}{D_{OL}}\left(1+\rm e_n \right),
	\end{eqnarray}
where $\theta_1^E$ represents the angular position of the outermost Einstein's ring. In \ref{ERfig}, we plot the angular position $\theta_1^E$ of the supermassive black holes SgrA$^*$ and M87. Likewise, we get the image magnification which mathematically represents the ratio of the solid angle formed due to the image and the source with the lens. For the $n$th image formed, the magnification is expressed as \cite{Bozza:2002zj, Bozza:2002af}
\begin{eqnarray}
\mu_n &=& \frac{1}{\beta} \Bigg[\frac{u_s}{D_{OL}}(1+\rm e_n) \Bigg(\frac{D_{OS}}{D_{LS}}\frac{u_s \rm e_n}{D_{OL}\Bar{a}}  \Bigg)\Bigg].
\end{eqnarray}
As the value $n$ increases, the magnification decreases and accordingly, the image becomes fainter. We have another important quantity, the angular separation $s$, which is the angular difference between the first and last image formed. Therefore, we have the important quantities of the relativistic images formed for a generic asymmetric rotating black hole as follows \cite{Bozza:2002zj}
\begin{eqnarray}
\theta_\infty &=& \frac{u_s}{D_{OL}},\\
s &=& \theta_1-\theta_\infty \approx \theta_\infty (\rm e^{\frac{\Bar{b}-2\pi}{\Bar{a}}}),\\
r_{\text{mag}} &=& \frac{\mu_1}{\sum{_{n=2}^\infty}\mu_n } \approx
e^{\frac{2 \pi}{\Bar{a}}}.
\end{eqnarray}
These observables are plotted in a realistic scenario for supermassive black holes such as Sgr A$^*$ and M87 \cite{Kormendy:2013} in \ref{fig:deflection_m87} and \ref{fig:magnification_m87}. The details of mass and other parameters for Sgr A$^*$ \cite{Do:2019vob} are: $M=4.3\times 10^6 M_{\odot}$ and $d=8.35$~Kpc, while for M87 \cite{Akiyama:2019eap}, $M=6.5 \times 10^9 M_{\odot} $ and $d=16.8$~Mpc. \\

Concerning the computations of various strong lensing observable parameters we follow the same procedures that are hugely applied to the four-dimensional black hole spacetimes in general as well as modified theories of gravity. All the observable parameters except the ones concerned with the Ohanian equation are well behaved and systematically evaluated. Regarding the five-dimensional lens equation, we simply employ the same techniques as it is done in the usual four-dimensional spacetimes \cite{Majumdar:2006zza, Chakraborty:2016lxo}. The justification for usage of such lens equations may lie on the following arguments--\\ 
\vspace{0.5mm}\\
i)~ We have rotations around the $\psi$ and $\phi$ directions, but when limited to the equatorial plane ($\theta=\pi/2$), the black hole rotation is constrained only in the $\psi$ direction. We cannot have rotation around the $\phi$ direction, which can be shown in the metric, \ref{final_metric}, where the only induced cross-term is $dt d\psi$. This leads us to conclude that the gravitational lensing by the five-dimensional rotating black hole is unusual when compared to the usual four-dimensional Kerr black hole. This is the first signature where we could hope that the extra dimensions must have imprint on the gravitational lensing effect. In addition, in the limit of vanishing rotation parameter ($j=0$), one can speculate that the function $H(r)=0$. It eventually leads to get the deflection angle of $\psi$ tending to zero, and subsequently \ref{final_metric} reduces to that of the usual five-dimensional Reissner-Nordstr$\Ddot{o}$m-AdS black hole spacetimes.\\
\vspace{0.5mm}\\
ii) Here we will limit our attention to investigate the deflection angle in the $\phi$ direction for the light rays passing close to the black hole in the equatorial plane since it can be observed by astronomical experiments. Moreover, it is very convenient for us to compare with the results obtained in the usual four-dimensional black hole spacetime.\\
\vspace{0.5mm}\\
iii) The expression for the angular position depends on the minimum impact parameter having a rigid base. The minimum impact parameter is derived when the effective potential is identically vanishing in all cases, regardless of the spacetime dimensions. In our understanding it could not have any effect in the derivation of the Ohanian equation. In the derivation of the angular position, the second term we encounter is the source-to-observer distance. It is the measured distance using the standard ruler. The parameters which might get affected in extra dimensions are the black hole parameters, not those measured using the standard rulers. The effects of extra dimensions are imprinted in the parameters like angular momentum, charge, and the mass. These parameters receive scaling corrections when transitioning from higher to lower dimensions or vice-versa. Even though the analytical results of the five-dimensional Ohanian equations (if possible to derive), it will not have significant deviations in its mathematical form. We could have the same mathematical expressions upto to some correction terms having not significant deviations. On this logical basis, we proceeded to calculate the observables using the four-dimensional Ohanian equation.\\
\vspace{0.5mm}\\
iv)~Other observable parameters like the angular separation, the image magnification and the relativistic time delay effects are systematically obtained in terms of the coefficients $\Bar{a}$ and $\Bar{b}$, which are obtained in the strong field region, when we approximate the deflection angle. These parameters are calculated independently from the lens equation. Here too, we see the effects of extra dimensions in the black hole parameters, but not in the mathematical derivation of the observable.
%%%%%%%%%%%%%%%%%%%%%%%%%%%%%%%%%%%%%%%%%%%%%%%%%%%%%%%%%%%%%%%%%%%%%%%%
\section{Time delay in the strong field limit}
\label{Time delay}
In this section, we derive the time required for the photon or light to go from $r_0$ to $r$ or from $r$ to $r_0$. We follow an approach similar to the one reported in the previous subsection for the deflection angle but with some subtraction strategies to treat the integrals. For more details, see Refs.~\cite{Bozza:2003cp, Hsieh:2021rru, Lu:2016gsf, Virbhadra:2007kw}.

For an observer at infinity, the time taken from the photon to travel from the source to the observer is given by
\begin{equation}
\label{time}
 T=\int_{t_{0}}^{t_{f}} \mathrm{dt}.
\end{equation}
Next, we change the integration variable from $t$ to $r$ and split the integral into two phases (approaching and leaving):
\begin{equation}
    \label{timesplit}
T=\int_{D_{L S}}^{r_{0}} \frac{\mathrm{dt}}{\mathrm{dr}}\mathrm{dr}+\int_{r_{0}}^{D_{O L}} \frac{\mathrm{dt}}{\mathrm{dr}} {\mathrm{dr}}.
\end{equation}

Utilizing the symmetry between approach and departure, we can combine the two integrals by extending the integration limits to infinity. This can be accomplished by subtracting two terms, given by
\begin{equation}
    T=2 \int_{r_{0}}^{\infty}\left|\frac{\mathrm{dt}}{\mathrm{dr}}\right| dr-\int_{D_{O L}}^{\infty}\left|\frac{\mathrm{dt}}{\mathrm{dr}}\right| \mathrm{dr}-\int_{D_{L S}}^{\infty}\left|\frac{\mathrm{dt}}{\mathrm{dr}}\right| \mathrm{dr}.
\end{equation}
If we consider two photons travelling on different trajectories, the time delay between them is
\begin{equation}
    \begin{split}\label{td2s}
      T_{1}-T_{2}=2 \int_{r_{0,1}}^{\infty}\left|\frac{\mathrm{dt}}{\mathrm{dr}}\left(r, r_{0,1}\right)\right| \mathrm{dr}-2 \int_{r_{0,2}}^{\infty}\left|\frac{\mathrm{dt}}{\mathrm{dr}}\left(r, r_{0,2}\right)\right| \mathrm{dr} \\
 -\int_{D_{O L}}^{\infty}\left|\frac{\mathrm{dt}}{\mathrm{dr}}\left(r, r_{0,1}\right)\right| \mathrm{dr}+\int_{D_{O L}}^{\infty}\left|\frac{\mathrm{dt}}{\mathrm{dr}}\left(r, r_{0,2}\right)\right| \mathrm{dr} \\
-\int_{D_{L S}}^{\infty}\left|\frac{\mathrm{dt}}{\mathrm{dr}}\left(r, r_{0,1}\right)\right| \mathrm{dr}+\int_{D_{L S}}^{\infty}\left|\frac{\mathrm{dt}}{\mathrm{dr}}\left(r, r_{0,2}\right)\right| \mathrm{dr}.
    \end{split}
\end{equation}

Supposing that the observer and the source are very far from the black hole, $\mathrm{dt} / \mathrm{dr}$ is effectively 1 in the last four integrals, which thus exactly cancel each other. We are thus left with the first two integrals.

Dividing \ref{coordeq} by \ref{radial}, we obtain
\begin{equation}
\label{dtdx}
    \frac{\mathrm{dt}}{\mathrm{dr}}=\tilde{P}_{1}\left(r, r_{0}\right) P_{2}\left(r, r_{0}\right),
\end{equation}
where
\begin{equation}
   \tilde{P}_{1}\left(r, r_{0}\right)=\sqrt{\frac{B(r)F(r_0)}{F(r)}},
\end{equation}
and $P_{2}$ is defined by \ref{fz}. It can be seen that $\mathrm{dt}/ \mathrm{dr}$ tends to one at large $r$ limit, and the two integrals in \ref{td2s} are separately divergent, while their difference is finite. The time delay results from the different paths the photons follow while they wind around the black hole. When the two photons are far from the black hole, $\mathrm{dt}/\mathrm{dr} \rightarrow 1$ and the two integrals compensate each other. Separating the two regimes, we can write individually convergent integrals. To achieve this, we subtract and add the function $\tilde{P}_{1}\left(r, r_{0, i}\right) / \sqrt{F_{0, i}}$ to each integrand. Supposing $r_{0,1}<r_{0,2}$, we can write
\begin{equation}
    \begin{split}
        \label{timediff}
        T_{1}-T_{2}=\tilde{T}\left(r_{0,1}\right)-\tilde{T}\left(r_{0,2}\right)+2 \int_{r_{0,1}}^{r_{0,2}} \frac{\tilde{P}_{1}\left(r, r_{0,1}\right)}{\sqrt{F_{0,1}}} \mathrm{dr}\\
 +2 \int_{r_{0,2}}^{\infty}\left[\frac{\tilde{P}_{1}\left(r, r_{0,1}\right)}{\sqrt{F_{0,1}}}-\frac{\tilde{P}_{1}\left(r, r_{0,2}\right)}{\sqrt{F_{0,2}}}\right] \mathrm{dr},
    \end{split}
\end{equation}
where
\begin{eqnarray}
    \label{Ttilde}
    \tilde{T}\left(r_{0}\right) &=& \int_{0}^{1} \tilde{R}\left(z, r_{0}\right) f\left(z, r_{0}\right) \mathrm{dz},\\
    \label{rtilde}
    \tilde{R}\left(z, x_{0}\right) &=& \frac{2r^2}{r_0} \tilde{P}_{1}\left(r, r_{0}\right)\left(1-\frac{1}{\sqrt{F_{0}} f\left(z, r_{0}\right)}\right),
\end{eqnarray}
and $f\left(z, r_{0}\right)$ is defined by \ref{fz}. We can verify that \ref{timediff} is similar to \ref{td2s} by substituting all the formulas, but now it is expressed as the sum of independently convergent integrals.

In actuality, the time taken by the light ray to wind around the black hole is represented by the integral $\tilde{T}\left(r_{0}\right)$. We have deducted a term from the definition of $R\left(z, r_{0}\right)$ that is negligible when the photon is near the black hole but cancels the integrand when the photon is distant from the black hole in order to cut off the integrands at large $r$ values. As we shall see in a moment, the residual terms of this subtraction are often subleading with respect to $\Delta \tilde{T}$ and are kept in the final two integrals of \ref{timediff}.

The integral~\ref{rtilde} can be solved following the same technique of the integral (\textcolor{blue}{38}), just replacing $R$ by $\tilde{R}$. The result is
\begin{equation}
    \label{Tu}
    \tilde{T}(u)=-\tilde{a} \ln \left(\frac{u}{u_{s}}-1\right)+\tilde{b}+O\left(u-u_{s}\right),
\end{equation}
where $u_{s}$ is defined by \ref{ab} and
\begin{equation}
    \label{atilde}
 \tilde{a}=\frac{\tilde{R}\left(0, r_{s}\right)}{2 \sqrt{q(r_{s})}},
 \end{equation}
 \begin{equation}
     \label{btilde}
     \tilde{b}=-\pi+\tilde{b}_{R}+\tilde{a} \ln\left(\frac{r^2_s[C^{\prime\prime}(r_s)F(r_s)-C(r_s)F^{\prime\prime}(r_s)]}{u_{s}F(r_s)\sqrt{F(r_s)C(r_s)}}\right)
 \end{equation}
with
\begin{equation}
    \label{bRtil}
    \tilde{b}_{R}=\int_{0}^{1}\left[\tilde{R}\left(z, r_{s}\right) f\left(z, r_{s}\right)-\tilde{R}\left(0, r_{s}\right) f_{0}\left(z, r_{s}\right)\right] \mathrm{dz}.
\end{equation}

\begin{figure}[h]
    \centering
    \includegraphics[scale=.6]{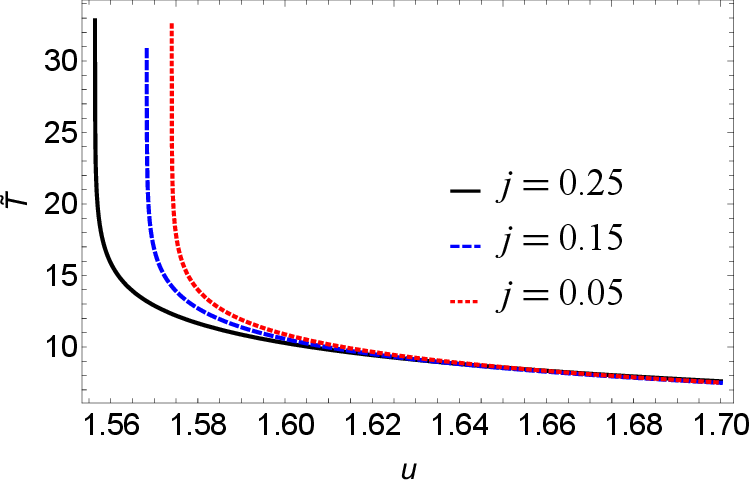}\hspace{0.1cm}
    \includegraphics[scale=.6]{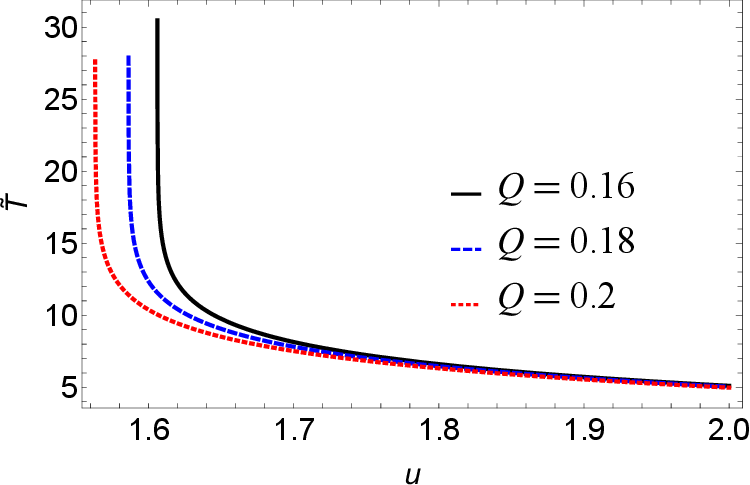}
    \caption{Plot of the variation of $\tilde{T}(u)$ with respect to the parameter $u$ for different values of the rotation parameter $j$ (the left plot) and constant value of the charge parameter $Q$ (the right plot).}
    \label{Tildet}
\end{figure}

\begin{figure}[h]
    \centering
    \includegraphics[scale=.6]{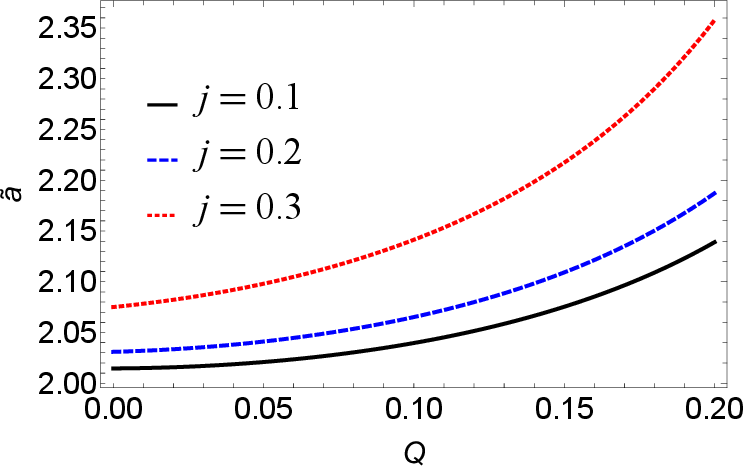}\hspace{0.1cm}
    \includegraphics[scale=.6]{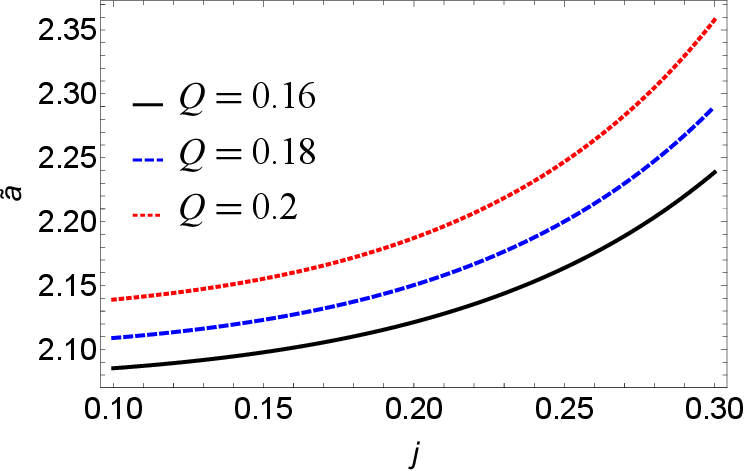}
       \caption{Plot of the variation of $\tilde{a}$ with respect to the charge parameter $Q$ (the left plot) and the rotation parameter $j$ (the right plot).}
    \label{atildefig}
\end{figure}

\begin{figure}[h]
    \centering
    \includegraphics[scale=.6]{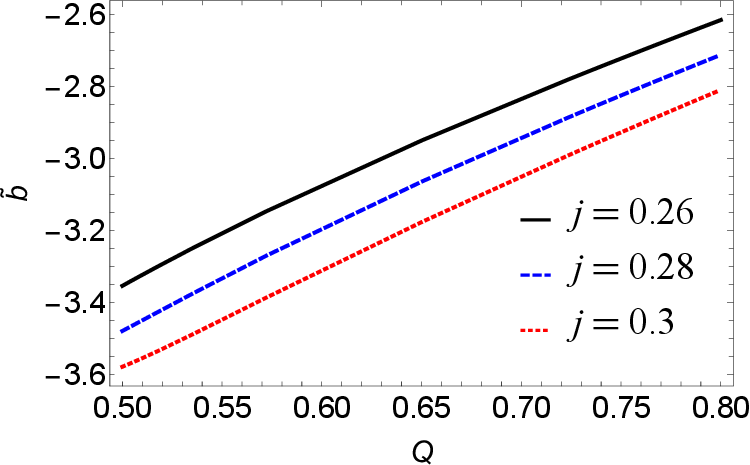}\hspace{0.1cm}
\includegraphics[scale=.6]{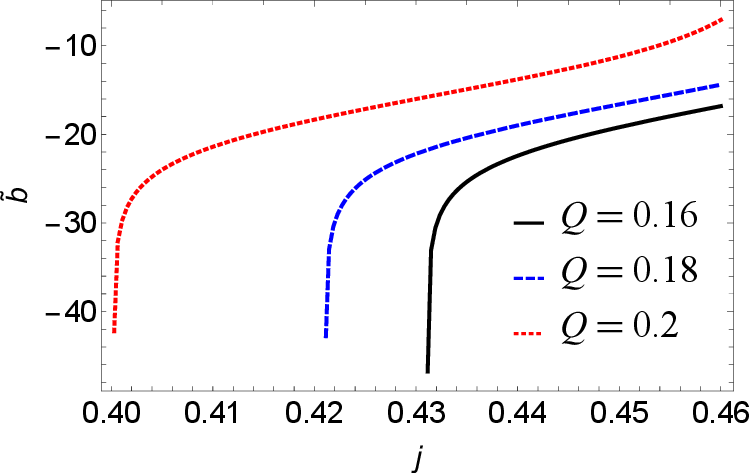}
    \caption{ Plot of the variation of $\tilde{b}$ with respect to the charge parameter $Q$ (the left plot) and the rotation parameter $j$ (the right plot).}
    \label{btildefig}
\end{figure}

We have shown the variation of $\tilde{T}(u)$ with respect to $u$ for different $j$ and $Q$ values in \ref{Tildet}.
Next, the variations of $\tilde{a}$ and $\tilde{b}$ with respect to the parameters $Q$ and $j$ are shown in \ref{atildefig} and \ref{btildefig}, respectively.

%%%%%%%%%%%%%%%%%%%%%%%%%%%%%%%%%%
\section{Conclusion}
%%%%%%%%%%%%%%%%%%%%%%%%%%%%%%%%%%
The strong gravitational lensing provides a very powerful tool to probe the nature of black holes in different modified theories of gravity. In addition, the extra dimension is a promising entity that helps us in determining a unified nature of the theory of everything.  We have investigated the strong gravitational lensing in the equatorial plane ($\theta=\pi/2$) of the five-dimensional charged equally rotating black hole in the presence of a negative cosmological constant. The presence of an additional factor of the $dt d\psi $ term in the spacetime metric itself, even for the equatorial plane, accounts for the extra dimension. We observed systematically the effects of the rotation and the charge parameters on the circular photon sphere radius for a fixed value of the curvature radius. The strong deflection angle and the coefficients $\bar{a}$, $\bar{b}$ were plotted in order to show the role of the black hole parameters. We found that the coefficient $\bar{a}$ increases with the charge parameter, while it decreases or increases with the rotation parameter $j$, depending upon the choice of the values of the charge parameter. However, the coefficient $\bar{b}$ increases with both the charge parameter and the rotation parameter. We also calculated the gravitational time delay effect and showed its variation with the charge parameter and the rotation parameter. In the near future, with the advent of new technologies, the detector EHT or square kilometre array may be operated to directly give the photon sphere radius of either the SgrA$^*$ or the M87 supermassive black hole. This may lead to putting constraints on the parameters of the five-dimensional, equally rotating black hole in the presence of a cosmological constant. This may help us fix which of the black hole solutions coming from the superstring or supergravity theory may be viable for the experimental extra dimensions to detect. At last but not least, we analysed the gravitational time delay effect by taking into consideration the observables accountable for the strong-field gravitational effects. These parameters are highly usable when investigating the quasinormal modes from the perspectives of the photon orbit radius and \textit{vis-\' a-vis} the minimum impact parameter.

Gravitational wave astronomy, through both ringdown or electromagnetic observation and the detection of the silhouettes of black holes, may provide us with information on the spacetime geometry near the black hole event horizon. The ringdown and the silhouettes are directly connected to a particular kind of null circular geodesics called the light rings \cite{Cardoso:2016rao}. For the four-dimensional case, in the asymptotically flat limit for a generic axisymmetric rotating black hole with spherical horizon topology near the event horizon, such light rings (at least one) are formed in each rotation sense \cite{Cunha:2020azh, Wu:2023eml}. For the five-dimensional case, there exists at least one light ring in the case of both the black hole and the naked singularity \cite{Tavlayan:2022hzl}. We hope to return to this issue for our concerned black hole in the near future. We also want to explore the timelike circular geodesics in the context of topological configurations for the massive particles, as they are important for accumulating particles around the accretion disk. Another future direction, in addition to the above analyses, should be devoted to the analysis of the quasinormal modes in correlation with the photon orbit radius and hence the lensing parameters.

\section*{Acknowledgement}
This work was supported by the National Natural Science Foundation of China (Grants No. 12475056, No.12347177, and No. 12247101), the 111 Project under (Grant No. B20063) and Lanzhou City's scientific research funding subsidy to Lanzhou University, the Gansu Province Major Scientific and Technological Special Project.

\bibliographystyle{cas-model2-names}

\bibliography{cas-refs}

\begin{thebibliography}{99}

\bibitem{LIGOScientific:2017ycc}
B.~P.~Abbott \textit{et al.} [LIGO Scientific and Virgo],
{\it GW170814: A Three-Detector Observation of Gravitational Waves from a Binary Black Hole Coalescence}, Phys. Rev. Lett. \textbf{119}, 141101 (2017).

\bibitem{LIGOScientific:2016aoc}
B.~P.~Abbott \textit{et al.} [LIGO Scientific and Virgo],
{\it Observation of Gravitational Waves from a Binary Black Hole Merger},
Phys. Rev. Lett. \textbf{116}, 061102 (2016).

\bibitem{LIGOScientific:2018dkp}
B.~P.~Abbott \textit{et al.} [LIGO Scientific and Virgo],
{\it Tests of General Relativity with GW170817},
Phys. Rev. Lett. \textbf{123}, 011102 (2019).

\bibitem{LIGOScientific:2019fpa}
B.~P.~Abbott \textit{et al.} [LIGO Scientific and Virgo],
{\it Tests of General Relativity with the Binary Black Hole Signals from the LIGO-Virgo Catalog GWTC-1}, Phys. Rev. D \textbf{100}, 104036 (2019).
\bibitem{Wielgus:2022heh}
M.~Wielgus, M.~Moscibrodzka, J.~Vos, Z.~Gelles, I.~Marti-Vidal, J.~Farah, N.~Marchili, C.~Goddi and H.~Messias,
{\it Orbital motion near Sagittarius A*-Constraints from polarimetric ALMA observations}, Astron. Astrophys. \textbf{665}, L6 (2022).
\bibitem{KAGRA:2023pio}
R.~Abbott \textit{et al.} [KAGRA, VIRGO and LIGO Scientific],
{\it Open Data from the Third Observing Run of LIGO, Virgo, KAGRA, and GEO},
Astrophys. J. Suppl. \textbf{267}, 29 (2023).

\bibitem{Akiyama:2019cqa}
K.~Akiyama {\it et al.},
{\it First M87 Event Horizon Telescope Results. I. The Shadow of the Supermassive Black Hole},
Astrophys.\ J.\  {\bf 875}, L1 (2019).

\bibitem{Akiyama:2019brx}
K.~Akiyama {\it et al.},
{\it First M87 Event Horizon Telescope Results. II. Array and Instrumentation},
Astrophys.\ J.\  {\bf 875}, L2 (2019).

\bibitem{Akiyama:2019sww}
K.~Akiyama {\it et al.},
{\it First M87 Event Horizon Telescope Results. III. Data Processing and Calibration},
Astrophys.\ J.\  {\bf 875}, L3 (2019).

\bibitem{Akiyama:2019bqs}
K.~Akiyama {\it et al.},
{\it First M87 Event Horizon Telescope Results. IV. Imaging the Central Supermassive Black Hole},
Astrophys.\ J.\  {\bf 875}, L4 (2019).

\bibitem{Akiyama:2019fyp}
K.~Akiyama {\it et al.},
{\it First M87 Event Horizon Telescope Results. V. Physical Origin of the Asymmetric Ring},
Astrophys.\ J.\  {\bf 875}, L5 (2019).

\bibitem{Akiyama:2019eap}
K.~Akiyama {\it et al.},
{\it First M87 Event Horizon Telescope Results. VI. The Shadow and Mass of the Central Black Hole},
Astrophys.\ J.\  {\bf 875}, L6 (2019).


\bibitem{Darwin}
C. Darwin, Proc. R. Soc. A \textbf{249}, 180 (1959).

\bibitem{Frittelli:1999yf}
S.~Frittelli, T.~P.~Kling and E.~T.~Newman,
{\it Space-time perspective of Schwarzschild lensing},
Phys. Rev. D \textbf{61}, 064021 (2000).


\bibitem{Eiroa:2010wm}
E.~F.~Eiroa and C.~M.~Sendra,
{\it Gravitational lensing by a regular black hole},
Class. Quant. Grav. \textbf{28}, 085008 (2011).

\bibitem{Bozza:2001xd}
V.~Bozza, S.~Capozziello, G.~Iovane and G.~Scarpetta,
{\it Strong field limit of black hole gravitational lensing},
Gen. Rel. Grav.\  {\bf 33}, 1535 (2001).

\bibitem{Bozza:2002zj}
V.~Bozza,
{\it Gravitational lensing in the strong field limit},
Phys. Rev. D {\bf 66}, 103001 (2002).

\bibitem{Bozza:2007gt}
V.~Bozza and G.~Scarpetta,
{\it Strong deflection limit of black hole gravitational lensing with arbitrary source distances},
Phys. Rev. D {\bf 76}, 083008 (2007).


\bibitem{Bozza:2002af}
V.~Bozza,
{\it Quasiequatorial gravitational lensing by spinning black holes in the strong field limit},
Phys. Rev. D \textbf{67}, 103006 (2003).


\bibitem{Bartelmann:2016dvf}
M.~Bartelmann and M.~Maturi,
{\it Weak gravitational lensing}, [arXiv:1612.06535 [astro-ph.CO]].


\bibitem{Hoekstra:2008db}
H.~Hoekstra and B.~Jain,
{\it Weak Gravitational Lensing and its Cosmological Applications},
Ann. Rev. Nucl. Part. Sci. \textbf{58}, 99 (2008).

\bibitem{Kuang:2022xjp}
X.~M.~Kuang and A.~\"Ovg\"un,
{\it Strong gravitational lensing and shadow constraint from M87* of slowly rotating Kerr-like black hole}, Annals Phys. \textbf{447}, 169147 (2022).


\bibitem{Sengo:2022jif}
I.~Sengo, P.~Cunha, V.P., C.~A.~R.~Herdeiro and E.~Radu,
{\it Kerr black holes with synchronised Proca hair: lensing, shadows and EHT constraints},
JCAP \textbf{01}, 047 (2023).

\bibitem{AbhishekChowdhuri:2023ekr}
A.~Chowdhuri, S.~Ghosh and A.~Bhattacharyya,
{\it A review on analytical studies in Gravitational Lensing},
Front. Phys. \textbf{11}, 1113909 (2023).

\bibitem{Ghosh:2022mka}
S.~Ghosh and A.~Bhattacharyya,
{\it Analytical study of gravitational lensing in Kerr-Newman black-bounce spacetime,}
JCAP \textbf{11}, 006 (2022)

\bibitem{EslamPanah:2020hoj}
B.~Eslam Panah, K.~Jafarzade and S.~H.~Hendi,
{\it Charged 4D Einstein-Gauss-Bonnet-AdS black holes: Shadow, energy emission, deflection angle and heat engine,}
Nucl. Phys. B \textbf{961}, 115269 (2020).


\bibitem{Hendi:2022qgi}
S.~H.~Hendi, K.~Jafarzade and B.~Eslam Panah,
{\it Black holes in dRGT massive gravity with the signature of EHT observations of M87*,}
JCAP \textbf{02}, 022 (2023).


\bibitem{Luminet:1979nyg}
J.~P.~Luminet,
{\it Image of a spherical black hole with thin accretion disk},
Astron. Astrophys. \textbf{75}, 228 (1979).

\bibitem{Ohanian:1987pc}
C. H. Ohanian,
{\it The black hole as a gravitational lensing}, American Journal of Physics. \textbf{55}, 428 (1987).

\bibitem{Virbhadra:1999nm}
K.~S.~Virbhadra and G.~F.~R.~Ellis,
{\it Schwarzschild black hole lensing},
Phys.\ Rev.\ D {\bf 62}, 084003 (2000).

\bibitem{Virbhadra:2002ju}
K.~S.~Virbhadra and G.~F.~R.~Ellis,
{\it Gravitational lensing by naked singularities},
Phys. Rev. D \textbf{65}, 103004 (2002).


\bibitem{Virbhadra:2008ws}
K.~S.~Virbhadra,
{\it Relativistic images of Schwarzschild black hole lensing,}
Phys. Rev. D \textbf{79}, 083004 (2009).


\bibitem{Virbhadra:1998dy}
K.~S.~Virbhadra, D.~Narasimha and S.~M.~Chitre,
{\it Role of the scalar field in gravitational lensing,}
Astron. Astrophys. \textbf{337}, 1 (1998).


\bibitem{Tsukamoto:2016jzh}
N.~Tsukamoto,
{\it Deflection angle in the strong deflection limit in a general asymptotically flat, static, spherically symmetric spacetime},
Phys. Rev. D \textbf{95}, 064035 (2017).

\bibitem{Tsukamoto:2017fxq}
N.~Tsukamoto,
{\it Black hole shadow in an asymptotically-flat, stationary, and axisymmetric spacetime: The Kerr-Newman and rotating regular black holes},
Phys. Rev. D \textbf{97}, 064021 (2018).

\bibitem{Islam:2021dyk}
S.~U.~Islam and S.~G.~Ghosh,
{\it Strong field gravitational lensing by hairy Kerr black holes},
Phys. Rev. D \textbf{103}, 124052 (2021).


\bibitem{Ghosh:2020spb}
S.~G.~Ghosh, R.~Kumar and S.~U.~Islam,
{\it Parameters estimation and strong gravitational lensing of nonsingular Kerr-Sen black holes},
JCAP \textbf{03}, 056 (2021).

\bibitem{Whisker:2004gq}
R.~Whisker,
{\it Strong gravitational lensing by braneworld black holes},
Phys. Rev. D \textbf{71}, 064004 (2005).

\bibitem{Abbas:2021whh}
G.~Abbas, A.~Mahmood and M.~Zubair,
{\it Strong deflection gravitational lensing for photon coupled to Weyl tensor in a charged Kiselev black hole},
Phys. Dark Univ. \textbf{31}, 100750 (2021).

\bibitem{Eiroa:2005ag}
E.~F.~Eiroa,
{\it Gravitational lensing by Einstein-Born-Infeld black holes},
Phys. Rev. D \textbf{73}, 043002 (2006).

\bibitem{Gyulchev:2006zg}
G.~N.~Gyulchev and S.~S.~Yazadjiev,
{\it Kerr-Sen dilaton-axion black hole lensing in the strong deflection limit},
Phys. Rev. D \textbf{75}, 023006 (2007).

\bibitem{Ghosh:2010uw}
T.~Ghosh and S.~Sengupta,
{\it Strong gravitational lensing across dilaton anti-de Sitter black hole},
Phys. Rev. D \textbf{81}, 044013 (2010).

\bibitem{Gyulchev:2012ty}
G.~N.~Gyulchev and I.~Z.~Stefanov,
{\it Gravitational Lensing by Phantom Black holes},
Phys. Rev. D \textbf{87}, 063005 (2013).

\bibitem{Molla:2023hog}
N.~U.~Molla, A.~Ali and U.~Debnath,
{\it Observational Signatures of Modified Bardeen Black Hole: Shadow and Strong Gravitational Lensing}, [arXiv:2307.11798 [gr-qc]].

\bibitem{Grespan:2023cpa}
M.~Grespan and M.~Biesiada,
{\it Strong Gravitational Lensing of Gravitational Waves: A Review},
Universe \textbf{9}, 200 (2023). 

\bibitem{Kumar:2022fqo}
J.~Kumar, S.~U.~Islam and S.~G.~Ghosh,
{\it Testing Strong Gravitational Lensing Effects of Supermassive Compact Objects with Regular Spacetimes}, Astrophys. J. \textbf{938}, 104 (2022).

\bibitem{KumarWalia:2022ddq}
R.~Kumar Walia,
{\it Observational predictions of LQG motivated polymerized black holes and constraints from Sgr A* and M87*}, JCAP \textbf{03}, 029 (2023).

\bibitem{Kumar:2021cyl}
J.~Kumar, S.~U.~Islam and S.~G.~Ghosh,
{\it Investigating strong gravitational lensing effects by supermassive black holes with Horndeski gravity}, Eur. Phys. J. C \textbf{82}, 443 (2022).

\bibitem{Lu:2021htd}
X.~Lu and Y.~Xie,
{\it Gravitational lensing by a quantum deformed Schwarzschild black hole},
Eur. Phys. J. C \textbf{81}, 627 (2021).


\bibitem{Ali:2021psk}
M.~S.~Ali and S.~Kauhsal,
{\it Gravitational lensing for stationary axisymmetric black holes in Eddington-inspired Born-Infeld gravity}, Phys. Rev. D \textbf{105},  024062 (2022).


\bibitem{Soares:2023err}
A.~R.~Soares, R.~L.~L.~Vit\'oria and C.~F.~S.~Pereira,
{\it Gravitational lensing in a topologically charged Eddington-inspired Born\textendash{}Infeld spacetime,}
Eur. Phys. J. C \textbf{83}, 903 (2023).


\bibitem{Hsieh:2021scb}
T.~Hsieh, D.~S.~Lee and C.~Y.~Lin,
{\it Strong gravitational lensing by Kerr and Kerr-Newman black holes},
Phys. Rev. D \textbf{103}, 104063 (2021).

\bibitem{Soares:2023uup}
A.~R.~Soares, C.~F.~S.~Pereira, R.~L.~L.~Vit\'oria and E.~M.~Rocha,
{\it Holonomy corrected Schwarzschild black hole lensing,}
Phys. Rev. D \textbf{108}, 124024 (2023)

\bibitem{Virbhadra:2022iiy}
K.~S.~Virbhadra,
{\it Distortions of images of Schwarzschild lensing,}
Phys. Rev. D \textbf{106}, 064038 (2022).

\bibitem{Virbhadra:2022ybp}
K.~S.~Virbhadra,
{\it Compactness of supermassive dark objects at galactic centers,}
[arXiv:2204.01792 [gr-qc]].


\bibitem{He:1999fe}
X.~G.~He, G.~C.~Joshi and B.~H.~J.~McKellar,
{\it Gravitational lensing and extra dimensions},
arXiv:hep-ph/9908469 [hep-ph].

\bibitem{Nandi:2024map}
K.~K.~Nandi, R.~N.~Izmailov, R.~K.~Karimov and A.~A.~Potapov,
{\it Observable strong field effects of extra spacetime dimension in the braneworld black hole},
Annals Phys. \textbf{470}, 169802 (2024).

\bibitem{Zahid:2024nvx}
M.~Zahid, F.~Sarikulov, C.~Shen, M.~Umaraliyev and J.~Rayimbaev,
{\it Shadow and quasinormal modes of novel charged rotating black hole in Born\textendash{}Infeld theory: Constraints from EHT results},
Phys. Dark Univ. \textbf{46}, 101616 (2024).

\bibitem{Shavelle:2024vwt}
K.~M.~Shavelle and D.~C.~M.~Palumbo,
{\it Prospects for the Detection of the Sgr A* Photon Ring with Next-generation Event Horizon Telescope Polarimetry},
Astrophys. J. Lett. \textbf{970}, L24 (2024).

\bibitem{Jiang:2023img}
H.~X.~Jiang, C.~Liu, I.~K.~Dihingia, Y.~Mizuno, H.~Xu, T.~Zhu and Q.~Wu,
{\it Shadows of loop quantum black holes: semi-analytical simulations of loop quantum gravity effects on Sagittarius~A* and M87*},
JCAP \textbf{01}, 059 (2024).

\bibitem{Ayzenberg:2023hfw}
D.~Ayzenberg, L.~Blackburn, R.~Brito, S.~Britzen, A.~Broderick, R.~Carballo-Rubio, V.~Cardoso, A.~Chael, K.~Chatterjee and Y.~Chen, \textit{et al.}
{\it Fundamental Physics Opportunities with the Next-Generation Event Horizon Telescope},
[arXiv:2312.02130 [astro-ph.HE]].


\bibitem{Shen:2006pa}
J.~Y.~Shen, B.~Wang and R.~K.~Su,
{\it The Signals from the brane-world black Hole},
Phys. Rev. D \textbf{74}, 044036 (2006).

\bibitem{Harris:2005jx}
C.~M.~Harris and P.~Kanti,
{\it Hawking radiation from a (4+n)-dimensional rotating black hole},
Phys. Lett. B \textbf{633}, 106 (2006).

\bibitem{Casals:2005sa}
M.~Casals, P.~Kanti and E.~Winstanley,
{\it Brane decay of a (4+n)-dimensional rotating black hole. II. Spin-1 particles},
JHEP \textbf{02}, 051 (2006).

\bibitem{Creek:2006ia}
S.~Creek, O.~Efthimiou, P.~Kanti and K.~Tamvakis,
{\it Graviton emission in the bulk from a higher-dimensional Schwarzschild black hole},
Phys. Lett. B \textbf{635}, 39 (2006).

\bibitem{Kanti:2005xa}
P.~Kanti and R.~A.~Konoplya,
{\it Quasi-normal modes of brane-localised standard model fields},
Phys. Rev. D \textbf{73}, 044002 (2006).

\bibitem{Kanti:2006ua}
P.~Kanti, R.~A.~Konoplya and A.~Zhidenko,
{\it Quasi-Normal Modes of Brane-Localised Standard Model Fields. II. Kerr Black Holes},
Phys. Rev. D \textbf{74}, 064008 (2006).

\bibitem{Creek:2006je}
S.~Creek, R.~Gregory, P.~Kanti and B.~Mistry,
{\it Braneworld stars and black holes},
Class. Quant. Grav. \textbf{23}, 6633 (2006).

\bibitem{Konoplya:2011qq}
R.~A.~Konoplya and A.~Zhidenko,
{\it Quasinormal modes of black holes: From astrophysics to string theory},
Rev. Mod. Phys. \textbf{83}, 793 (2011).

\bibitem{Kodama:2009rq}
H.~Kodama, R.~A.~Konoplya and A.~Zhidenko,
 {\it Gravitational instability of simply rotating AdS black holes in higher dimensions},
Phys. Rev. D \textbf{79}, 044003 (2009).

\bibitem{Konoplya:2008rq}
R.~A.~Konoplya and A.~Zhidenko,
{\it Stability of higher dimensional Reissner-Nordstrom-anti-de Sitter black holes},
Phys. Rev. D \textbf{78}, 104017 (2008).

\bibitem{Nozawa:2008wf}
M.~Nozawa and T.~Kobayashi,
{\it Quasinormal modes of black holes localized on the Randall-Sundrum 2-brane},
Phys. Rev. D \textbf{78}, 064006 (2008).

\bibitem{Chen:2007ay}
S.~Chen, B.~Wang, R.~K.~Su and W.~Y.~P.~Hwang,
{\it Greybody factors for rotating black holes on codimension-2 branes},
JHEP \textbf{03}, 019 (2008).

\bibitem{Chen:2007pu}
S.~Chen, B.~Wang and R.~K.~Su,
{\it Hawking radiation in a rotating Kaluza-Klein black hole with squashed horizons},
Phys. Rev. D \textbf{77}, 024039 (2008).

\bibitem{Liu:2010wh}
Y.~Liu, S.~Chen and J.~Jing,
{\it Strong gravitational lensing in a squashed Kaluza-Klein black hole spacetime},
Phys. Rev. D \textbf{81}, 124017 (2010).

\bibitem{Sadeghi:2012bj}
J.~Sadeghi, A.~Banijamali and H.~Vaez,
{\it Strong Gravitational Lensing in a Charged Squashed Kaluza- Klein Black hole},
Astrophys. Space Sci. \textbf{343}, 559 (2013).

\bibitem{Ji:2013xua}
L.~Ji, S.~Chen and J.~Jing,
{\it Strong gravitational lensing in a rotating Kaluza-Klein black hole with squashed horizons}, JHEP \textbf{03}, 089 (2014).

\bibitem{Chen:2011ef}
S.~Chen, Y.~Liu and J.~Jing,
{\it Strong gravitational lensing in a squashed Kaluza-Klein G\"odel black hole},
Phys. Rev. D \textbf{83}, 124019 (2011).
Chen:2011ef, Sadeghi:2013ssa
\bibitem{Sadeghi:2013ssa}
J.~Sadeghi, J.~Naji and H.~Vaez,
{\it Strong gravitational lensing in a charged squashed Kaluza-Klein G\"odel black hole},
Phys. Lett. B \textbf{728}, 170 (2014).


\bibitem{Majumdar:2006zza}
A.~S.~Majumdar and N.~Mukherjee,
{\it Gravitational lensing by higher dimensional black holes},
Contributed to 11th Marcel Grossmann Meeting on General Relativity, 1707.

\bibitem{Chakraborty:2016lxo}
S.~Chakraborty and S.~SenGupta,
{\it Strong gravitational lensing A probe for extra dimensions and Kalb-Ramond field},
JCAP \textbf{07}, 045 (2017).

\bibitem{Markeviciute:2018cqs}
J.~Markeviciute,
{\it Rotating Hairy Black Holes in AdS$_5\times$S$^5$},
JHEP \textbf{03}, 110 (2019).

\bibitem{Cvetic:2004hs}
M.~Cvetic, H.~Lu and C.~N.~Pope,
{\it Charged Kerr-de Sitter black holes in five dimensions},
Phys. Lett. B \textbf{598}, 273 (2004).

\bibitem{Raffaelli:2021gzh}
B.~Raffaelli,
{\it Hidden conformal symmetry on the black hole photon sphere},
JHEP \textbf{03}, 125 (2022).

\bibitem{Hadar:2022xag}
S.~Hadar, D.~Kapec, A.~Lupsasca and A.~Strominger,
{\it Holography of the photon ring},
Class. Quant. Grav. \textbf{39}, 215001 (2022).

\bibitem{Kapec:2022dvc}
D.~Kapec, A.~Lupsasca and A.~Strominger,
{\it Photon rings around warped black holes},
Class. Quant. Grav. \textbf{40}, 095006 (2023).

\bibitem{Chen:2023zvd}
Y.~Chen, W.~Guo, K.~Shi and H.~Zhang,
{\it SL(2, R) \texttimes{} U(1) symmetry and quasinormal modes in the self-dual warped AdS black hole},
JHEP \textbf{06}, 075 (2023).

\bibitem{Chen:2022fpl}
B.~Chen, Y.~Hou and Z.~Hu,
{\it On emergent conformal symmetry near the photon ring},
JHEP \textbf{05}, 115 (2023).

\bibitem{Fransen:2023eqj}
K.~Fransen,
{\it Quasinormal modes from Penrose limits},
Class. Quant. Grav. \textbf{40}, 205004 (2023)

\bibitem{Chan:1996yk}
J.~S.~F.~Chan and R.~B.~Mann,
{\it Scalar wave falloff in asymptotically anti-de Sitter backgrounds},
Phys. Rev. D \textbf{55}, 7546 (1997).

\bibitem{KalyanaRama:1999zj}
S.~Kalyana Rama and B.~Sathiapalan,
{\it On the role of chaos in the AdS/CFT connection},
Mod. Phys. Lett. A \textbf{14}, 2635 (1999).

\bibitem{Horowitz:1999jd}
G.~T.~Horowitz and V.~E.~Hubeny,
{\it Quasinormal modes of AdS black holes and the approach to thermal equilibrium},
Phys. Rev. D \textbf{62}, 024027 (2000).

\bibitem{Hashimoto:2019jmw}
K.~Hashimoto, S.~Kinoshita and K.~Murata,
{\it Einstein Rings in Holography},
Phys. Rev. Lett. \textbf{123}, 031602 (2019).

\bibitem{Hashimoto:2018okj}
K.~Hashimoto, S.~Kinoshita and K.~Murata,
{\it Imaging black holes through the AdS/CFT correspondence},
Phys. Rev. D \textbf{101}, 066018 (2020).

\bibitem{Bhattacharyya:2010yg}
S.~Bhattacharyya, S.~Minwalla and K.~Papadodimas,
{\it Small Hairy Black Holes in $AdS_5\times S^5$},
JHEP \textbf{11}, 035 (2011).


\bibitem{Frolov:2003en}
V.~P.~Frolov and D.~Stojkovic,
{\it Particle and light motion in a space-time of a five-dimensional rotating black hole},
Phys. Rev. D \textbf{68}, 064011 (2003).

\bibitem{Frolov:2002xf}
V.~P.~Frolov and D.~Stojkovic,
{\it Quantum radiation from a five-dimensional rotating black hole},
Phys. Rev. D \textbf{67}, 084004 (2003).


\bibitem{Gavazzi:2008}
R.~Gavazzi, T.~Treu, L. V. E. Koopmans, A. S. Bolton, L. A. Moustakas, S. Burles, P. J. Marshall, Astrophys. J., \textbf{677}, 1046 (2008).


\bibitem{Kormendy:2013}
J.~Kormendy and L.~C.~Ho,
{\it Coevolution (Or Not) of Supermassive Black Holes and Host Galaxies},
Ann. Rev. Astron. Astrophys. \textbf{51}, 511 (2013).


\bibitem{Do:2019vob}
T. Do et al.,{\it Unprecedented variability of Sgr A* in NIR}, Science \textbf{365}, 664 (2019).

\bibitem{Bozza:2003cp}
V.~Bozza and L.~Mancini,
{\it Time delay in black hole gravitational lensing as a distance estimator},
Gen. Rel. Grav. \textbf{36}, 435 (2004).

\bibitem{Hsieh:2021rru}
T.~Hsieh, D.~S.~Lee and C.~Y.~Lin,
{\it Gravitational time delay effects by Kerr and Kerr-Newman black holes in strong field limits},
Phys. Rev. D \textbf{104}, 104013 (2021).


\bibitem{Lu:2016gsf}
X.~Lu, F.~W.~Yang and Y.~Xie,
{\it Strong gravitational field time delay for photons coupled to Weyl tensor in a Schwarzschild black hole},
Eur. Phys. J. C \textbf{76}, 357 (2016).

\bibitem{Virbhadra:2007kw}
K.~S.~Virbhadra and C.~R.~Keeton,
{\it Time delay and magnification centroid due to gravitational lensing by black holes and naked singularities,}
Phys. Rev. D \textbf{77}, 124014 (2008).


\bibitem{Cardoso:2016rao}
V.~Cardoso, E.~Franzin and P.~Pani,
{\it Is the gravitational-wave ringdown a probe of the event horizon?},''
Phys. Rev. Lett. \textbf{116} (2016), 171101, Erratum: Phys. Rev. Lett. \textbf{117}, 089902 (2016).

\bibitem{Cunha:2020azh}
P.~V.~P.~Cunha and C.~A.~R.~Herdeiro,
{\it Stationary black holes and light rings},
Phys. Rev. Lett. \textbf{124}, 181101 (2020).

\bibitem{Wu:2023eml}
S.~P.~Wu and S.~W.~Wei,
{\it Topology of light rings for extremal and non-extremal Kerr-Newman Taub-NUT black holes without $\mathbb{Z}_2$ symmetry},''
[arXiv:2307.14003 [gr-qc]].

\bibitem{Tavlayan:2022hzl}
A.~Tavlayan and B.~Tekin,
{\it Light rings around five-dimensional stationary black holes and naked singularities},''
Phys. Rev. D \textbf{107}, 024016 (2023). 


\end{thebibliography}

%%%%%%%%%%%%%%%%%%%%%%%%%%%%%%%%%%%%%

\end{document}